%% file: vanzella.tex
\documentclass[12pt,preprint]{aastex}

\begin{document}

\slugcomment {To appear in the Nov.~2001 issue of {\it The
Astronomical Journal}}

\shorttitle {Multicolor observation of HDF-S}
\shortauthors {E.~Vanzella et al.}

\title {Multicolor observations of the Hubble Deep Field South
\footnote{Based on
observations with the NASA/ESA \textit{Hubble Space Telescope},
and on observations collected at the \textit{ESO-VLT}
as part of the programme 164.O-0612}
}

\author {Eros~Vanzella\altaffilmark{2,3}, 
Stefano Cristiani\altaffilmark{4,3},
Paolo Saracco\altaffilmark{5},
Stephane Arnouts\altaffilmark{2}, 
Simone Bianchi\altaffilmark{2},
Sandro D'Odorico\altaffilmark{2},
Adriano Fontana\altaffilmark{6},
Emanuele Giallongo\altaffilmark{6},
Andrea Grazian\altaffilmark{2,3}
}

\altaffiltext{2}{ESO - European Southern Observatory,
Karl-Schwarzschild-Str. 2, D-85748 Garching bei M\"unchen, Germany}
\altaffiltext{3}{Dipartimento di Astronomia, Universit\'a di Padova,
vicolo dell'Osservatorio 2, I-35122 Padova, Italy}
\altaffiltext{4}{Space Telescope European Coordinating Facility,
Karl-Schwarzschild-Str. 2, D--85748 Garching, Germany}
\altaffiltext{5}{Osservatorio Astronomico di Brera, via E. Bianchi 46,
Merate, Italy}
\altaffiltext{6}{Osservatorio Astronomico di Roma, via
dell'Osservatorio 2, Monteporzio, Italy}

\begin{abstract}

We present a deep multicolor ($U$, $B$, $V$, $I$, $Js$, $H$, $Ks$)
catalog of galaxies in the Hubble Deep Field South, based on
observations obtained with the HST WFPC2 in 1998 and VLT-ISAAC in 1999.
The photometric procedures were tuned to derive a catalog
optimized for the estimation of photometric redshifts. In particular we
adopted a ``conservative'' detection threshold which resulted in
a list of 1611 objects.

The behavior of the observed source counts is in general agreement with the
result of \citet{casertano00} in the Hubble Deep Field South and
\citet{williams96} in the Hubble Deep Field North, while the
corresponding counts in the Hubble Deep Field North provided by
\citet{soto99} are systematically lower by a factor 1.5
beyond $I_{AB}$=26.
After correcting for the incompleteness of the source counts, the object
surface density at $I_{AB} \le 27.5$ is estimated to be $220$ per
square arcmin, in agreement with the corresponding measure of
\citet{volonteri00}, and providing an estimate of the Extragalactic
Background Light in the $I$ band consistent with the work of \citet{madau00}.

The comparison between the median $V-I$ color in the Hubble Deep Field North
and South shows a significant difference around $I_{AB} \sim 26$,
possibly due to the presence of large scale structure at $z\sim 1$
in the HDF-N.

High-redshift galaxy candidates (90 $U$ dropout and 17 $B$
dropout) were selected by means of color diagrams, down to
a magnitude $I_{AB} = 27$, with a surface density 
of $(21 \pm 1)$ and $(3.9 \pm 0.9)$ 
per square arcmin, respectively. 

Eleven extremely red objects (with $(I-K)_{AB} > 2.7$) were selected
down to $K_{AB}=24$, plus three objects whose upper limit to the
Ks flux is still compatible with the selection criterion.
The corresponding surface density of EROs is  $(2.5\pm0.8)$ per square
arcmin ($(3.2\pm 0.9)$ per square arcmin if we include the three $Ks$
upper limits).
They show a remarkably non-uniform spatial distribution and are
classified with roughly equal fractions in the categories of
elliptical and starburst galaxies.
\end{abstract}
 

\section {Introduction}

The Hubble Deep Field South (hereafter HDF-S) consists of a large set
of observations of an otherwise unremarkable field around the QSO
J2233-606 (z $=$ 2.24), taken in parallel by three instruments aboard
the Hubble Space Telescope (HST): the Wide Field and Planetary Camera
2 (WFPC2), the Space Telescope Imaging Spectrograph (STIS) and NICMOS.

The observations are described in a series of papers: \citet{williams00}
provide details on the field selection and the overall
strategy, \citet{fruchter01} the NICMOS observations, \citet{gardner00}
the STIS imaging observations, \citet{savaglio99} the
STIS spectroscopic observations, and \citet{casertano00} the WFPC2
observations (the main field).

The WFPC2 observations were centered around $\alpha = 22^h \ 32^m \
56.22^s$, $\delta = -60^{\circ} \ 33' \ 02.7''$ (J2000) and consist of
a total of approximately 450 Ks worth of exposures in the four bands
F300, F450, F606, F814 (in the following indicated as $U$, $B$, $V$, 
and $I$, respectively), like the original HDF \citep{williams96}.

The HDFs represent an important part of the frontier studies of the
distant universe.  They have revolutionized the understanding of high
redshift galaxies, providing resolved images of faint objects, and have
contributed in a wide variety of ways to shaping the debate over
issues such as the origin of elliptical galaxies and the importance of
obscured star formation.  The addition of deep near infrared images to
the database provided by the WFPC2 is essential to monitor the SEDs of
the objects on a wide baseline and address a number of key issues such as
the total stellar content of baryonic mass, the effects of dust
extinction, the dependence of morphology on the rest frame wavelength,
the photometric redshifts, and the detection and nature of extremely red
objects (EROs).

For these reasons deep NIR imaging of the HDF-S was carried out
with VLT-ISAAC and the present paper describes a new catalog built
from the combined WFPC2 - ISAAC database.

\section {The Observations}

\subsection {Optical:
WFPC2}\label{sec:scheduling_strategy}

In Table~\ref{table:tobs}, we summarize the observations in the
filters $U~B~V~I$ and list the estimated magnitude limits
at 10-sigma within apertures of 2$\times$FWHM for the combined frames
(more details are given in Casertano et al. 2000).  The optical deep
images of WFPC2 are available at the URL:
$ftp://archive.stsci.edu/pub/hdf{\_}south/version1/wfpc2{\_}version2/$.

\subsection {Infrared: ISAAC}

The IR data were obtained with the ISAAC infrared
imager/spectrometer \citep{moorwood99} at the ESO VLT-UT1
telescope. ISAAC is equipped with a 1024 $\times$ 1024 pixel Rockwell
Hawaii array providing a pixel scale of 0.147 arcsec/pix and a total
field of view of about 2.5 $\times$ 2.5 arcmin.  The observed field is
centered at $\alpha = 22^h \ 32^m \ 55^s$, $-60^{\circ}33^{'}08^{''}$.
The observations were carried out over several nights from September
to December 1999 under homogeneous seeing conditions: about 0.6
arcsec. The HDF-S was imaged in the $Js$, $H$, $Ks$ bands: in
Table~\ref{table:tobsIR} we summarize the number of frames and the
relevant total exposure time per band.  
The $Js$ and $Ks$ filters were adopted instead of the standard
$J$ and $K$ because: 1) the standard J filter of ISAAC (with
central wavelength 1.25 $\mu m$ and width 0.29 $\mu m$) has leaks in
the K band and the atmosphere defines the red edge of the filter. With
the $Js$ filter it is possible to obtain more accurate photometry.
2) The $Ks$ filter, having a shorter cutoff at long wavelength, makes
it possible to reduce the thermal background and increase the S/N
ratio. Further details on the instrument are available in the ISAAC
user manual (Cuby, Lidman and Moutou, 2001,
$http://www.eso.org/instruments/isaac/index.html$).

The photometric calibration of the observations was carried out by
observing several standard stars from the list of the Infrared NICMOS
Standard Stars \citep{persson98} with magnitudes ranging between
10 and 12.

The reduction and co-adding were performed using the facilities
available at the Merate Observatory.  Raw frames, corrected for
flat-field, bias, dark current and thermal noise were processed
with DIMSUM
\footnote[1]{A package of IRAF scripts by Eisenhardt, Dickinson and
Ward, available at ftp://iraf.noao.edu/contrib/dimsumV2}
(Deep Infrared Mosaicking Software) to produce the sky-subtracted
frames; for each exposure a sky background image is generated as the
median of a set (from 3 to 6) of time adjacent frames in which sources
were previously masked out.

The sky-subtracted frames were inspected and sky residuals were
removed where present, using the IRAF task $imsurfit$. 
The images were rescaled to the same median value by adding to
each of them a suitable constant term.
Finally, shifting and co-adding were performed with DIMSUM. 
The final
co-added image is the average of the shifted frames.
The infrared data analyzed in this work were presented in a
paper of \citet{saracco01} to which we refer for further details
on the data reduction.

In  Table~\ref{table:tobsIR}, we summarize 
the observations and list the observed magnitude limits at 5-sigma within
apertures of 2$\times$FWHM of the combined frames. 
The total time exposures in the filters $Js$, $H$, $Ks$ are
$7^{h}00^{m}$, $6^{h}02^{m}$ and $8^{h}07^{m}$  respectively.


\section {Data Reduction}

Figure~\ref{fig:schema_riduz} shows the global scheme of the
reduction. 

\subsection {Object Detection}
The catalog was produced using the Source Extractor package
(SExtractor Bertin \& Arnouts 1996) in a slightly modified and more
flexible version (see below).

Source detection and deblending were carried out on the
inverse-variance-weighted sum of the $V$ and $I$ drizzled
images. While other techniques for weighting are possible (e.g. Szalay
et al. 1999), for most sources this summed image provides the maximum
limiting depth. The resulting source position and isophotes are then
used for subsequent photometric analysis in each of the individual
bands. The combined $V+I$ image is significantly deeper than any
of the individual images, and using it to define the object catalog
allows, in practice, the detection of all the objects with ``normal''
SEDs that are detectable in the individual optical bands.

Source detection with SExtractor follows a standard ``connected
pixel'' algorithm.  To identify sources, the detection image is first
convolved with a fixed smoothing kernel, for which we use a circular
Gaussian with FWHM = 2.5 pixels.  Pixels with convolved values higher
than a fixed threshold are marked as potential sources. This threshold
is in unit of $\sigma_{sky}$, where $\sigma_{sky}$ as a function of
position comes from the background RMS map produced by SExtractor. 
The variation of S/N as a
function of position is thus automatically taken into account.

After thresholding, regions consisting of more than a certain number
of contiguous pixels are counted as sources. The aim of the present
work is to provide a high-quality photometric catalog, sampling the
SEDs of the sources from the UV to the NIR, in order to produce
reliable photometric redshifts.  Therefore we used a
``conservative'' detection strategy: the source detection threshold
was set to 1.26 $\sigma_{sky}$
(per pixel in the convolved image) and the minimum
area to 16 drizzled pixels, i.e. 0.025 square arcsec.  After
extensive tests of SExtractor it
became clear that no single set of input parameters provides
the desired detection depth for isolated objects and faint companions
of bright galaxies, and the required rejection of spurious sources in
correspondence with stellar diffraction spikes 
\footnote{In the reduction seven relatively bright stars were masked 
using the Weight Watcher package.}.
This is a standard problem with SExtractor and at the moment an ideal
solution does not exist.
For example, \citet{casertano00}
ran SExtractor two times with different detection thresholds. Objects 
from the deeper catalog are selectively removed and substituted with
the corresponding ones from the shallower catalog.
In this way, however, it is not easy to achieve consistent estimates of
the fluxes (and in general also of the colors) with the different
thresholds.
Moreover the use of different detection thresholds can lead to the
loss of objects near bright sources.

In order to obtain the desired detection depth and at the same time 
the required rejection of spurious sources with a single value of the
detection threshold we introduced a modification of SExtractor
that makes  the detection strategy more flexible.
To solve particular cases of clearly wrong detection (as
described above), a table is prepared in which it is possible to
define different values of the parameters in different regions of the
frame (in the present application the $deblend$\_$mincont$,
the minimum contrast parameter for deblending, and the
$detect$\_$minarea$, the minimum number of pixels above threshold
triggering detection \footnote{For details on the parameters used by
SExtractor see the SExtractor user guide at
$ftp://ftp.iap.fr/pub/from\_users/bertin/sextractor/$.}
).  
Outside the specified subframes the program
uses the $global$ parameters defined in the standard configuration
file. In our case the global values are: $detect$\_$minarea$ = 16
drizzled pixels and $deblend$\_$mincont$ = 0.02. Therefore in a single
run we obtained a unique catalog where all objects are detected
over the same threshold ($detect$\_$thresh$, the detection threshold).
We defined 78 rectangular subframes, 57 with $deblend$\_$mincont$
ranging from 0. to 1. (maximum deblending and no deblending) and
the others with suitable $detect$\_$minarea$ basically to avoid
spurious detections.
It is also possible to personalize both parameters at the same time in
a given subframe, and define more subframes one inside the other.

Although such a procedure is in direct contrast to the convenience 
of a fully automated algorithm able to
process a large amount of data, we had to resort to it in order to
ensure the highest quality of the result.  In
Figures~\ref{fig:sex1},\ref{fig:sex2} we show two examples of
correction through the definition of subframes. The accurate detection
is also helpful for the correct estimation of the object coordinates
(X,Y) on the $V+I$ image. The correct centering of the photometric
apertures is in fact a critical issue for the estimation of the colors,
when SExtractor is used in $double$ $image$ $mode$ (see below).

The final result of the detection procedure is a list of 1611 objects.

The remaining process to obtain the photometry followed the standard
SExtractor package. 

\subsection {Magnitudes}

\subsubsection{Optical catalog}

To estimate the total magnitudes ($f(I)_{tot}$) of the
objects, we used a procedure described by \citet{djorgovski95}. 
For brighter objects, where the isophotal area (detected in
$V$+$I$ image) is larger than 0.497 square arcsec (corresponding to an
aperture of 20 pixels in diameter), we used the SExtractor $best$
magnitude as defined in the SExtractor manual. The $best$ magnitude is
provided by an adaptive aperture method (mag $auto$),
except if a neighbor is
suspected to bias the measurement by more than 0.1 mag. 
In the latter case,
the corrected isophotal magnitude is taken \citep{SEx}.

For objects with an isophotal area smaller than 0.497 square arcsec, a
magnitude estimated in a circular aperture of 0.8 arcsec
(corresponding roughly to 5 FWHM) was used, corrected to the total
$best$ magnitude provided by SExtractor, assuming that the wings
follow a stellar profile. The corrections turned out to be independent
of the magnitude and in terms of fluxes correspond to 1.061, 1.012,
1.010, 1.012 in the $U$, $B$, $V$ and $I$ band, respectively.  
There are 231 objects (out of 1611) with isophotal area greater than
0.497 square arcsec. The flux $best$ was computed (by SExtractor)
for 208 objects out of 231 as a flux $auto$ and for the remaining
23 as isophotal corrected.
These 23 objects are smoothly distributed in the range $22< I_{AB} <
25$ and the RMS difference with respect to the flux $auto$ is
0.07 magnitudes.

We selected the $I$ as the fundamental band, and we referred to it
for the determination of the object colors. If not otherwise
specified, magnitudes are expressed in the AB system, that is a system
based on a spectrum which is flat in $f_{\nu}$:
$m=-2.5Logf_{\nu}+48.59$ \citep{oke74}.

In order to apply the photometric redshift codes it is mandatory to obtain a
reliable estimation of the colors for each object. The colors were
measured using the same fixed aperture (0.8 arcsec) in the different
optical filters.

For the $U$, $B$, $V$ bands the flux is obtained by 
first estimating the ratio of the flux in each band with
respect to the $I$ band within the same fixed aperture and then
rescaling it to the total $I$ flux: 
\begin{equation}
F(n) = [{f(n) \over f(I)}]_{fix. aperture} \times f(I)_{tot}
\end{equation}
with $n$ = $U$, $B$, $V$. The corresponding error is estimated by
standard propagation.

We ran SExtractor in the so-called $double$ $image$ $mode$: the $V$+$I$
image was used for the detection of the sources and the images in the
bands $U$, $B$, $V$, $I$ for the photometric measurements.

\subsubsection{Infrared catalog}
The infrared frames have different orientation, pixel size and Point
Spread Function (PSF) with respect to the HST images.
In order to determine the color indices between the optical 
(in particular the $I$ band) 
and the near infrared bands ($Js$, $H$, $Ks$) we adopted the
following strategy:
\begin{enumerate}
\item{The IR images were rebinned to the same pixel size and 
orientation as the WFPC2 images.}
\item{To estimate colors, the $I$ frame was smoothed in turn to
the same effective PSF of each of the images $Js$, $H$, $Ks$
(seeing $\simeq$ 0.6 arcsec) using a Fast Fourier Transform (FFT)
approach.

In order to obtain the correct transfer function, we extracted
the median stellar PSF (obtained from 13 stellar objects) 
from each IR image and from the $I$ image, 
and we transformed them in the frequency
space through a Fast Fourier Transform.  

The division of the FFTs of the infrared PSFs ($Js$, $H$, $Ks$) 
by the FFT of the $I$-band PSF
provided the transformed transfer functions (i.e. the
Fourier transform of the kernel).  For each band, we
inverse-transformed it in order to obtain the shape of the filter to be
applied to the $I$ image (in the coordinates domain).  The noise due
to the presence of the aliasing frequencies in the Fourier space was
reduced with a suitable filter.  The flux is preserved.

At the end we obtained three $I$-band smoothed images with the
``same'' PSF as the infrared images, namely: $I_{j}$, $I_{h}$,
$I_{k}$.}

\item{We ran SExtractor in $double$ $image$ $mode$ with the detection
in the $V$+$I$ image, and the photometry on the three smoothed $I$
band images ($I_{j}$, $I_{h}$, $I_{k}$) and on the three infrared
images $Js$, $H$, $Ks$. We then computed the $I_{j}-Js$,
$I_{h}-H$ and $I_{k}-Ks$ using a fixed aperture of $2 \times FWHM$
diameter.}

\end{enumerate}

The estimated flux for each IR band is, again:
\begin{equation}
F(n) = [{f(n) \over f(In)}]_{fix. aperture} \times f(I)_{tot}
\end{equation}
where $n$ runs over $Js$, $H$, $Ks$, and
$[{f(n) \over f(In)}]_{fix. aperture}$ is
the ratio of the flux in each IR band with
respect to the $I$ smoothed image
within the same fixed aperture.
The $f(I)_{tot}$ is the total $I$ magnitude obtained as described in
Sect. 3.2.1.

SExtractor was used (in $double$ $image$ $mode$) to carry out the
photometry on all the objects detected in the $V+I$ image.  The
fraction of isolated objects that are not significantly detected in
the IR images can be identified by comparing their fluxes with the
corresponding errors.

There are however 317 sources (but only 7 with $K_{AB}<24$)
for which a spurious IR flux is generated by contamination 
from a nearby brighter object. Those
cases were identified on the basis of the optical images (using
the $f(I)_{tot} / f(I_{n})$ ratio) and flagged as upper limits (see
the description of the flag $f_{we}$ below).

\subsection{Photometric errors for the infrared images}
Because neighboring pixels are strongly correlated, the noise in the
ISAAC images is not well described by the background RMS map produced 
by SExtractor, resulting in unrealistic estimates of the photometric
uncertainties.  In order to determine
the correction factor to the errors provided by SExtractor we
adopted the following strategy.

Synthetic stellar objects with known magnitude were generated with
a relatively high signal to noise ratio (S/N = 20) and
randomly inserted in 35 different relatively isolated regions of
the frame. Their fluxes were then measured again with SExtractor. The
dispersion in magnitude provides a reliable estimate of the real
error which was used to rescale the formal SExtractor errors by a
constant factor (3.25 in the three bands Js, H ,Ks).

\section{The Catalog}
The catalog\footnote{ 
The full photometric catalog with a description file is
available in fits format at the following URL:
http://www.stecf.org/hstprogrammes/ISAAC/ISAAC.html}
is presented in Table~\ref{table:catalog}.

For each object in the catalog we report the following parameters:

{\bf ID}: The identification number. The objects in the list are sorted 
by increasing right ascension.

{\bf x, y}: Abscissa and ordinate in pixels of the objects in the
version 2 WFPC2 images.

{\bf RA, DEC}: Coordinates RA and DEC (J2000) in the version 2 WFPC2 images.

{\bf$m_{i}$ and $\sigma(m_{i})$}: The magnitude in the $I$ image and
its uncertainty. It is the $best$ SExtractor magnitude if the
isophotal area of the object is greater then 0.497 square arcsec
(314 pixel), otherwise it is the fixed aperture magnitude (diameter
0.8 arcsec).
  
{\bf flux} and {\bf $\sigma$ (flux)}: The aperture fluxes, with their 
uncertainties (for the IR images they were estimated according to 
the procedure described in Sect. 3.2.2).

{\bf s/g}: The star-galaxy separation provided by SExtractor applied
to the $V$+$I$ image to which the best PSF corresponds. This
classifier gives output values in the range (0-1) (0 for galaxies and
1 for stars) and represents the degree of membership of an object to
one of the two classes.

{\bf $f_{s}$}: The internal flag provided by SExtractor (see Bertin \&
Arnouts 1996). It is a short integer containing, coded in decimal, all
the extraction flags as a sum of powers of two: 
\begin{list}
{ }
\item{ 1 - The object has neighbors, bright and close enough to
significantly bias the photometry, or bad pixel.}
\item{ 2 - The object was originally blended with another one.}
\item{ 4 - At least one pixel of the object is saturated (or very close to).}
\item{ 8 - The object is truncated (too close to an image boundary).}
\item{ 16 - Object's aperture data are incomplete or corrupted.}
\item{ 32 - Object's isophotal data are incomplete or corrupted.}
\item{ 64 - A memory overflow occurred during deblending.}
\item{128 - A memory overflow occurred during extraction.}
\end{list}

{\bf $f_{we}$}: Flag for infrared fluxes, for the eventual
contamination of the flux estimation from neighboring objects. In
these cases the flux measurement is considered as an upper limit.
The coding is a sum of powers of 2 (ranging from 0 to
7): for example 7 corresponds to 111 in binary and means that the
infrared fluxes in $Js$, $H$, $Ks$ are all upper limits, 4 (100) only
the $Js$ band is an upper limit, 3 (011) both $H$ and $Ks$ bands are
upper limits, 0 (000) all fluxes in the infrared bands have been
measured without problems, etc.

\section{Discussion}

\subsection{Source Counts}
 
Figure~\ref{fig:CountsIab} shows the behavior of the observed source counts: 
the result of the present work is consistent with the corresponding
estimates
in the HDF-S of \citet{casertano00} down to $I_{AB}$ $\sim$ 28
and with the counts in the HDF-N by \citet{williams96}.
On the other hand, if we compare our source counts with the result 
in the HDF-N by Fern\'andez-Soto et
al. (1999), beyond $I_{AB}$ $\sim$ 26 
the HDF-N counts turn out to be systematically lower with
respect to the HDF-S by about a factor 1.5,
while the source counts in the HDF-S by \citet{volonteri00} appear to be
higher in this range by about the same factor. These discrepancies are
to be ascribed to differences in the approach used to carry out the
photometry, in particular the profile fitting method for Fern\'andez-Soto et
al. (1999) and different parameters for the
source detection in the case of \citet{volonteri00}.

In order to compute the incompleteness of the source counts 
two different approaches were followed,
with the aim of obtaining 
an estimate of the systematic uncertainties.
In the first method 46 galaxies in the magnitude
interval $24.5 < I_{AB}< 25$ were selected and their extracted $V+I$
images were randomly inserted in the $V+I$ WF image, suitably
dimmed by increasing factors in order to cover the magnitude range
down to $I_{AB}=28$. Then the source detection was performed with 
the same SExtractor parameters used in the original image.
For each 0.5 magnitude bin 30 simulations were carried out.
The resulting incompleteness is listed in Col.3 ($c_{gal}$) of
Table~\ref{table:Counts}.
For the magnitude bins brighter than $I_{AB}=24.5$ no simulations were 
carried out and the incompleteness factor corresponds to the masked area
due to the brightest objects in the field (see Sect.3.1).

The same approach was also applied using a reference sample of 46
stellar images, in order to estimate a lower limit of the
incompleteness.  The result is listed in Col.4 ($c_{star}$) of
Table~\ref{table:Counts} and is increasingly different from the case
of the galaxy simulations beyond $I_{AB}=26.5$.
 
The estimated surface density, corrected by the
incompleteness factor for galaxies, is listed in Col.5 ($Nc_{gal}$) 
and is in very
good agreement with the corrected source counts of \citet{volonteri00}.

Fig.~\ref{fig:EBL} shows the extragalactic background light (EBL) per
half magnitude bin,
$i_\nu=10^{-0.4(I_{\rm AB}+48.6)}N(I_{\rm AB})$, as a function of 
$I$ magnitude.
The agreement with the work of \citet{madau00} is very good down to
$I_{AB} \simeq 26$. Beyond this limit the \citet{madau00} values are
close to our estimates corrected by the incompleteness factor for
stars rather than for galaxies.
In any case the uncertainty of this correction has little
effect on the total amount of EBL in $I$.
Adopting the values of \citet{madau00} down to $I_{AB}=22$ and
our counts at fainter magnitudes,
the EBL in the $I$ band turns out to be
$8.7$ and $8.8$ nW m$^{-2}$ sr$^{-1}$ using the incompleteness
correction for stars and galaxies, respectively.
These values are well within the uncertainty of the estimate by
\citet{madau00}, $8.04_{-0.92}^{+1.62}$ nW m$^{-2}$ sr$^{-1}$, which
arises mostly from field-to-field variations in the number of 
relatively bright galaxies.

\subsection{Colors}
In Figure~\ref{fig:Dcasert} we show the comparison between the optical 
colors of our catalog as a function of the
$I_{AB}$ magnitude and those obtained
by \citet{casertano00}.  The objects were selected with the
magnitude $U$, $B$, $V$, $I$ brighter than 27.0, 27.5, 27.5, 27.5
respectively.  The main differences (in particular in the $U$ band)
are due to the different strategy adopted in terms of detection and
photometry: we computed the colors using a fixed aperture, while
\citet{casertano00} computed the isophotal colors.

The comparison between the colors of the HDF-S (our catalog) and HDF-N
(Fern\'andez-Soto et al. 1999) is shown in
Figure~\ref{fig:colorV_I}. There is a significant difference in the
median color $V-I$ between the HDF-S and the HDF-N around magnitude
$I_{AB}$ $\sim$ 26. This could be due to systematic differences in the
distributions of the galaxy redshifts in the two fields, in particular
to the presence of large scale structure at $z\sim 1$ \citep{cohen00} 
and will be discussed in a forthcoming paper.  
Beyond $I_{AB}$=28 the limited
depth of the V image (29.5, 2$\sigma$) gives rise to systematic
effects in the color $V$-$I$.

Color criteria, which are sensitive to the presence
of a Lyman continuum break superposed on an
otherwise flat UV spectrum, have been shown to successfully
identify high redshift star-forming galaxies (e.g. Madau et al. 1996).
In Figure~\ref{fig:CandU} we show the selection of $U$-band dropouts
with $I_{AB} < 27$ (according to the criteria defined by Madau et al.
1996).  The dashed lines outline the selection region within which 
candidate $2.5 < z < 3.5$ objects are identified.  Objects undetected in
the $U$ band (with signal-to-noise $<$ 1) are plotted as triangles at the
1$\sigma$ limits to their $(U-B)_{AB}$ colors. We adopted a
similar technique to select candidate $3.5 < z < 4.5$ galaxies with
$I_{AB}< 27$ (see Figure~\ref{fig:CandB}). The selection area is
outlined with dashed lines. The undetected objects in the $B$ band (with
signal-to-noise $<$ 1) are plotted as triangles. Symbol sizes scale
with the $I_{AB}$ magnitude of the object.

90 $U$-band dropouts and 17 $B$-band dropouts are identified,
equivalent to a surface density of $(21 \pm 1)$ and $(3.9 \pm 0.9)$ 
per square arcmin, respectively. 
They are listed in Table~\ref{table:Udropouts} and
Table~\ref{table:Bdropouts}.

\subsection{Extremely Red Objects}
The selection of Extremely Red Objects (EROs) is shown in 
Figure~\ref{fig:CandEROs} (upper panel).
EROs are defined as sources with $(I-K)_{AB} > 2.7$, 
corresponding  to the color of
passively evolving elliptical galaxies at $z > 1$.
The selection was carried out down to $K_{AB} < 24$.
Table~\ref{table:TABEROs} lists the ID number and coordinates
of 11 EROs in the field (plus 3 more objects whose upper limit to the
Ks flux is still compatible with the selection criterion).

%

\citet{saracco01}, using the same infrared images as the present work, 
selected 20 EROs 
on the basis of a color criterion $Js - Ks > 2.4$, 
corresponding to a surface density of 2.7 per square arcmin.
Ten of these EROs lie in the area covered by our multicolor catalog
and six of them were selected with the above-described 
$(I-K)_{AB} > 2.7$ threshold.
In the catalog of \citet{saracco01} there is only one object 
detected with a S/N$>5$ in the $Ks$ band that
is not present in our catalog.

This object is clearly seen in the $Ks$ image, while
no optical counterpart is present in the $I$ band.
It is marked in Figure~\ref{fig:CandEROs} with an upward-pointing
arrow as a 5$\sigma$ lower limit at $I_{AB}$$\sim$28 and listed in
Table~\ref{table:TABEROs} with the ID 9999.  
Adding this object to our sample the
EROs surface density turns out to be $(3.2\pm 0.9)$ per square arcmin, 
$(2.5\pm0.8)$ per square arcmin without the 3 upper limits in the Ks band.

The EROs surface density measured in the present work is similar to
the result obtained by \citet{saracco01} in the same HDF-S field
(but over a larger area).  On the other hand the surface density
observed by \citet{saracco01} in the Chandra Deep Field
\citep[CDF]{CDF} is significantly different ($1.2 \pm 0.5$ per
square arcmin).  The discrepancy between the two fields is probably
due to the clustering properties of EROs associated to the relatively
small field of CDF and HDF-S. \cite{daddi00}
detect a strong clustering signal of EROs, selected
on the basis of the $R-K$ color from a $K_{\rm Vega}$ $\sim$ 19 limited sample
in a wide area of $\sim$ 25 square arcmin.  We select EROs on a sample
3 magnitudes fainter and with different color criteria. Thus, the
results of \cite{daddi00} cannot be directly applied to our data.
Nevertheless, the distribution of EROs in the HDF-S field is not
uniform: there are 10 EROs
(2 of them are upper limits in the Ks band) 
inside the upper WFPC2 chip. 
   
In the EROs sample there is an interesting radio source identified and  
discussed by \citet{norris99} which has an $(I-Ks)_{AB}$= 3.45.
This object is unusual in that its radio-optical ratio is about 1000
times higher than that of any galaxy known in the local Universe.

We adopted the photometric criterion of \citet{pozzetti00} to
distinguish between the two EROs populations of ellipticals and
starbursts.
In doing so we assume that our EROs sample belongs to the redshift
range of applicability of the criterion, $1<z<2$.
While an old stellar population is characterized by sharp breaks
(especially around 4000\AA), young dusty galaxies show shallower SEDs
in the rest-frame optical part of the spectrum because dust extinction
does not produce sharp breaks.
It is possible to measure the steepness of the SED for objects with
$1<z<2$ using the observed J-K color.
In the lower panel of the Figure~\ref{fig:CandEROs} the I-K vs. J-K
plane for the ERO sample is shown. The dashed line separating the
elliptical from the starburst galaxies corresponds to:
$J_{AB}-K_{AB}= 0.36(I_{AB} - K_{AB})+0.045$.


The morphological selection of early-type galaxies is 
usually carried out by fitting 
a generalized de~Vaucouleurs law 
($\mu_r-\mu_0\propto (r/r_e)^{1/n}$ to the luminosity profile of the objects,
where $\mu_r$ is the surface brightness at the radius $r$,
$r_e$ is the half light radius and $n$ is 
the Sersic index \citep{sersic68}. For elliptical galaxies $n >2$
(see, for example, \citet{rodighiero01}).
In the present case, the EROs have typically
$I_{AB}>26$ and except for the few brighter objects it is impossible 
to compute a reliable fit, even on the $V+I$ image. 
For the fainter EROs a rough morphological classification was
carried out comparing the central intensity vs. the area of the images.
A visual inspection of all the EROs was also carried out to check 
the results.
In Table~\ref{table:TABEROs} the EROs are listed, according to their
morphology, as early type
($E$, 3 objects), late type ($L$, 1 object) 
and objects too faint for a reliable classification ($S$, 11 objects).
In the lower panel of Figure~\ref{fig:CandEROs}, the three ellipticals
are marked with a cross inside the symbol, and the late-type galaxy is 
shown with a filled symbol. 
The other points belong to the (S) category.
The classification obtained with the morphological method agrees with
the photometric classification for the three ellipticals.
The object morphologically classified as ``late-type'' corresponds to an
IR upper limit and it is not clear whether it can be considered an ERO.

In the upper panel of Figure~\ref{fig:CandEROs}, the 
elliptical galaxies (according to the photometric classification)
are marked with a star. About half of the EROs in the sample seem to belong to
the ``Starburst'' category and, in general, show an $I-K$ color 
redder than the corresponding $I-K$ of the ``Ellipticals''.

In Figures~\ref{fig:EROsPOSTER1} and \ref{fig:EROsPOSTER2} the
$V+I$ and Ks images of the ERO sample are shown.

It is worth noting that the above-mentioned radio galaxy (marked with
a triangle in Figure~\ref{fig:CandEROs}) belongs to the ``early-type'' 
class according to the color-color separation
technique.  \citet{rodighiero01} also put it in a sample
of ``early-type'' objects, estimating for this
object a Sersic index $n \sim 2.2$
(G. Fasano, private communication) 
i.e. at the border of the early-/late-type discrimination.

\acknowledgments

$Acknowledgments.$ We are greatly indebted to F.Poli for valuable
suggestions about the FFT method to measure colors from images of
different PSF and to Pamela Bristow for her careful reading of the 
manuscript. 
This work was partially supported by the ASI
grants under contract number ARS-98-226 and ARS-96-176,
by the research contract of the University of Padova ``The High
redshift Universe: from HST and VLT to NGST''
and by the Research Training Network "The Physics of the Intergalactic
Medium" set up by the European Community under the contract
HPRN-CT2000-00126 RG29185. EV acknowledges the European Southern
Observatory, the Space Telescope European Coordinating Facility
and the Merate Observatory for their hospitality 
during the visits when this work was completed.

\clearpage

\input vanzella.figures.tex
\input vanzella.tab.tex

\end{document}

%% file: vanzella.figures.tex
\begin{figure}
\plotone{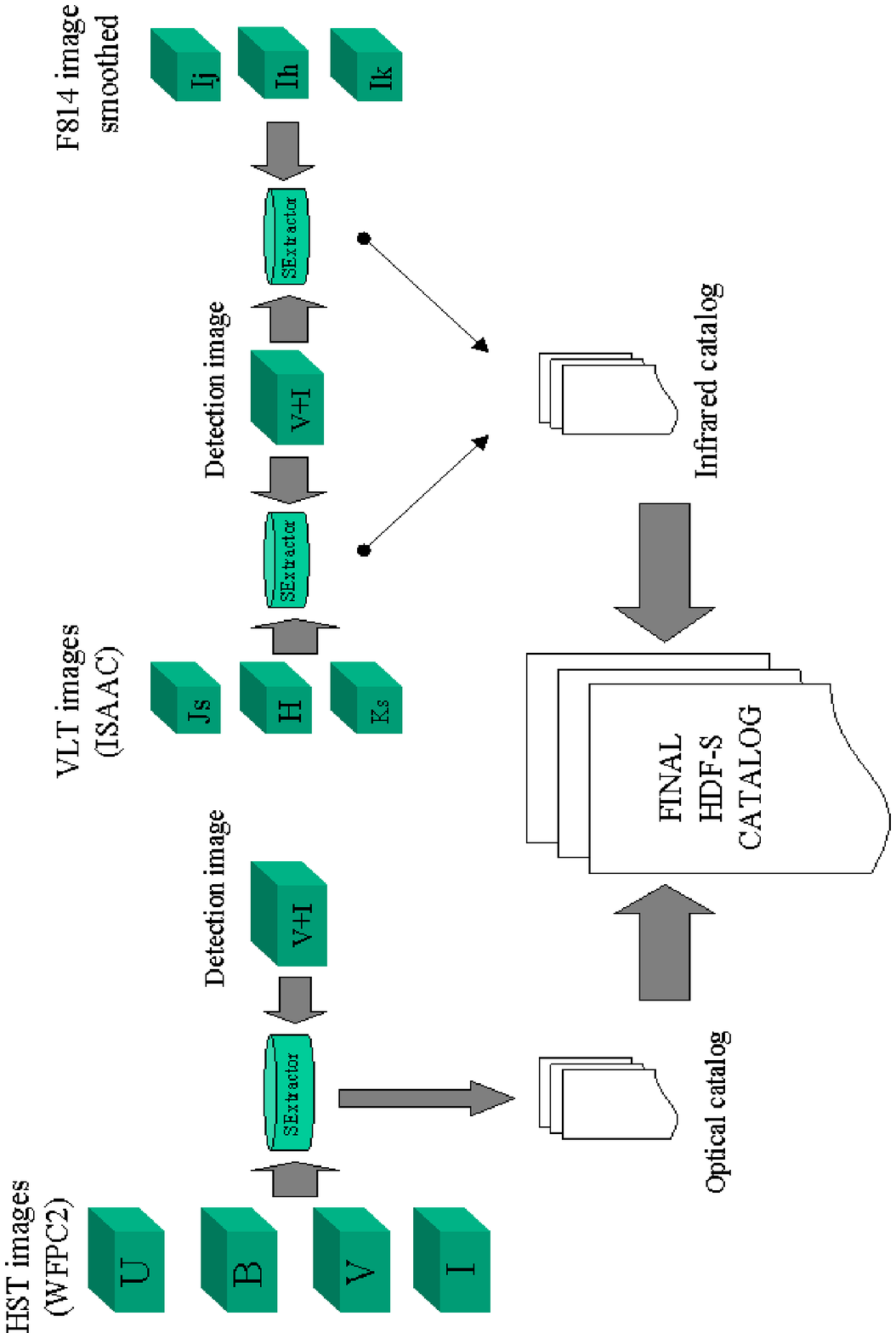}
 \caption[vanzella.FIG1.ps] {Global scheme of reduction.
 \label{fig:schema_riduz}}
\end{figure}

\clearpage 
\begin{figure}
\plottwo{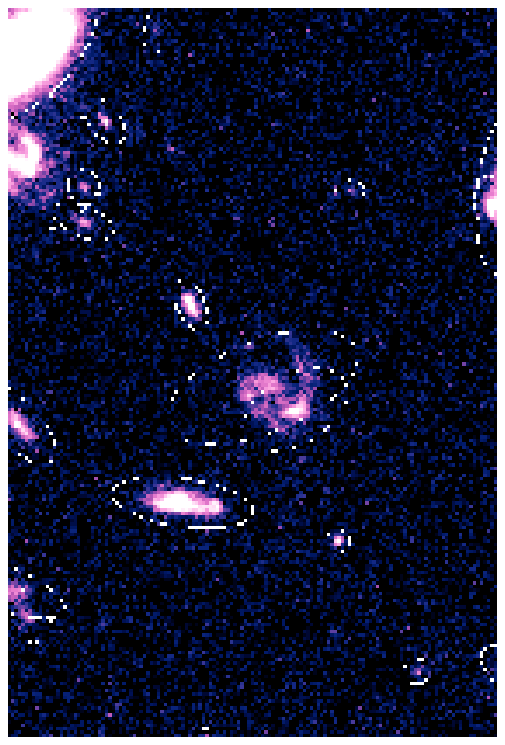}{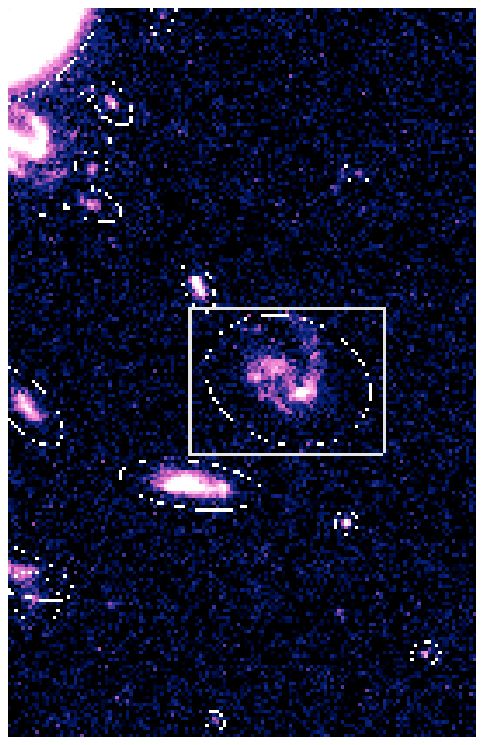}
 \caption{Left panel, detection without subframes,
 global parameter $deblend$\_$mincont$=0.02. In the right panel
 detection with modified version of SExtractor, inside the rectangle
 $deblend$\_$mincont$=1.0.\label{fig:sex1}}
  \end{figure}

 \begin{figure}
\plottwo{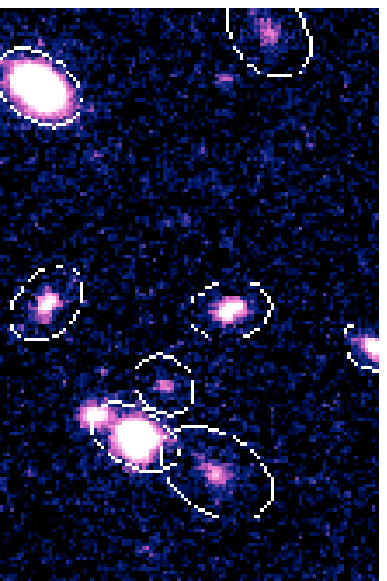}{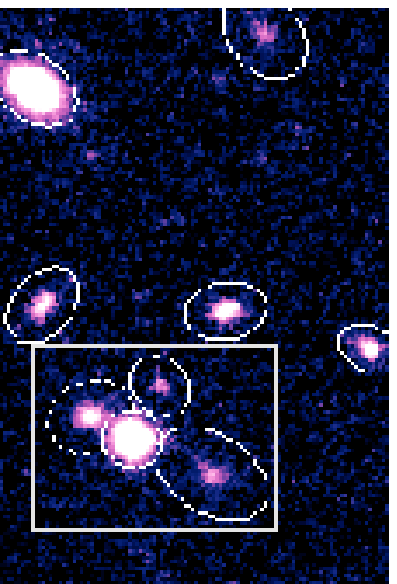}
 \caption{Left panel detection without subframes,
 global parameters $deblend$\_$mincont$=0.02.  In the right panel
 detection with modified version of SExtractor, inside the rectangle
 $deblend$\_$mincont$=0.0004 (the coordinates X,Y of brighter object
 split are more accurate).
\label{fig:sex2}}
  \end{figure}

 \begin{figure}
 \plotone{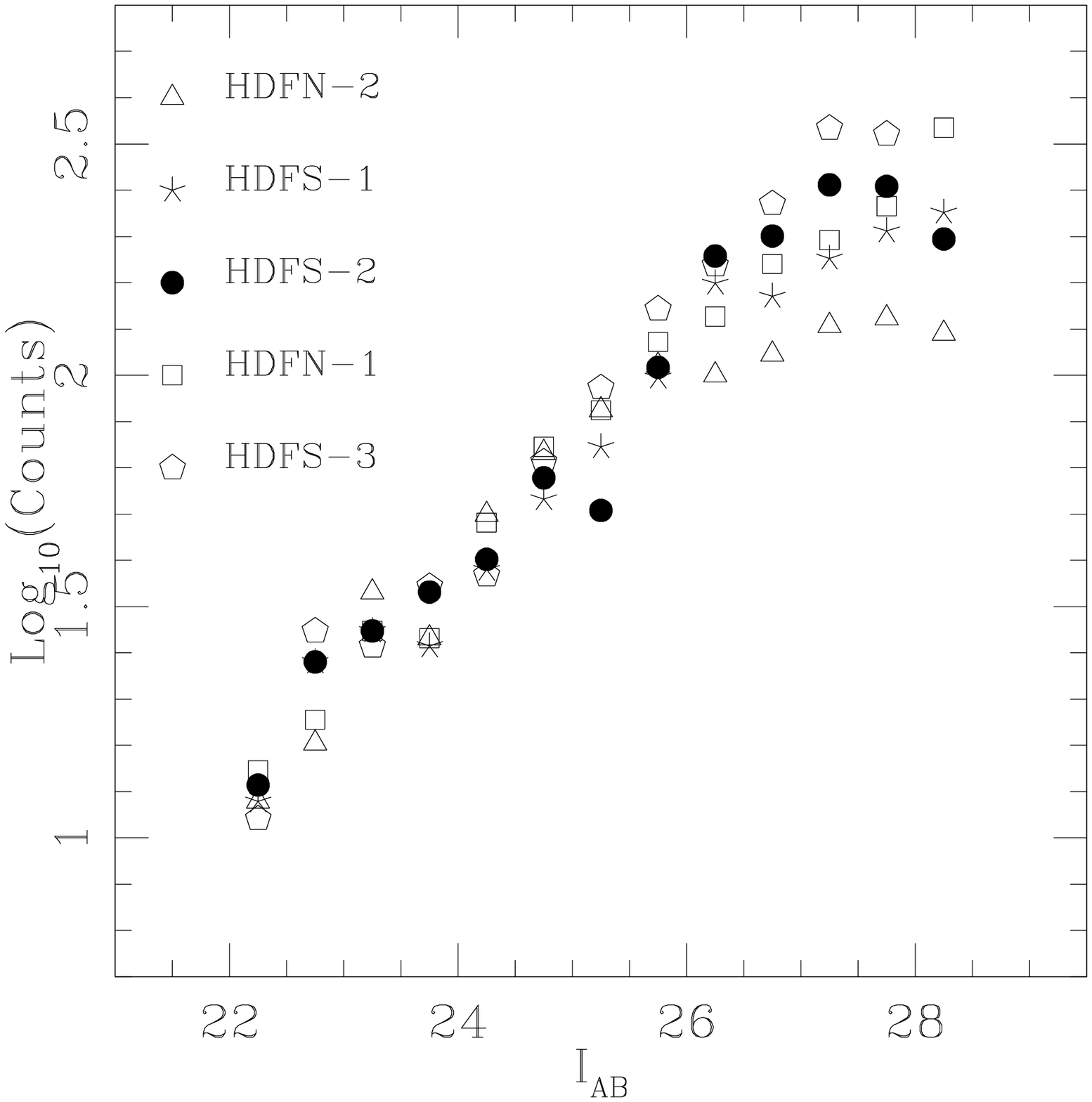}
 \caption{Comparison between the object counts
 in the HDF-S and the HDF-N. The present catalog (HDF-S-2),
 \citet{casertano00} (HDF-S-1), \citet{volonteri00} (HDF-S-3); \citet{soto99}
 (HDF-N-2) and \citet{williams96}
 (HDF-N-1). All the samples, except the \citet{volonteri00}, 
 include the planetary camera.
 \label{fig:CountsIab}}
  \end{figure}

 \begin{figure}
 \plotone{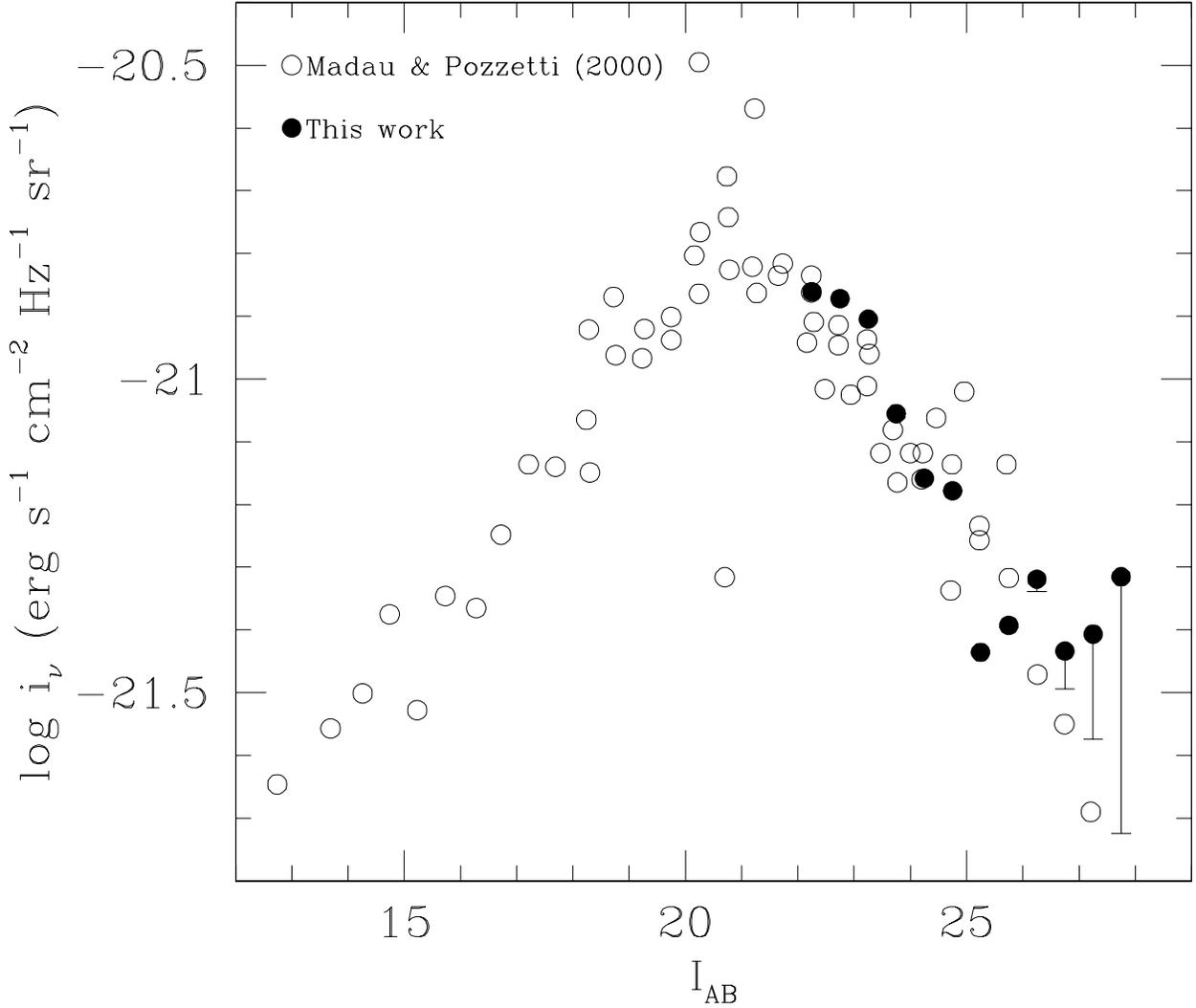}
 \caption{The extragalactic background light per
half magnitude bin,
$i_\nu=10^{-0.4(I_{\rm AB}+48.6)}N(I_{\rm AB})$, as a function of 
$I$ magnitude. The open circles are taken from \citet{madau00}.
The filled circles were obtained in the present work adopting the
incompleteness correction estimated for galaxies (see text). The error 
bars at $I_{AB} \simeq 26$ show the lower limit corresponding to the
incompleteness correction estimated for objects with a stellar PSF.
\label{fig:EBL}}
  \end{figure}

 \begin{figure}
 \plotone{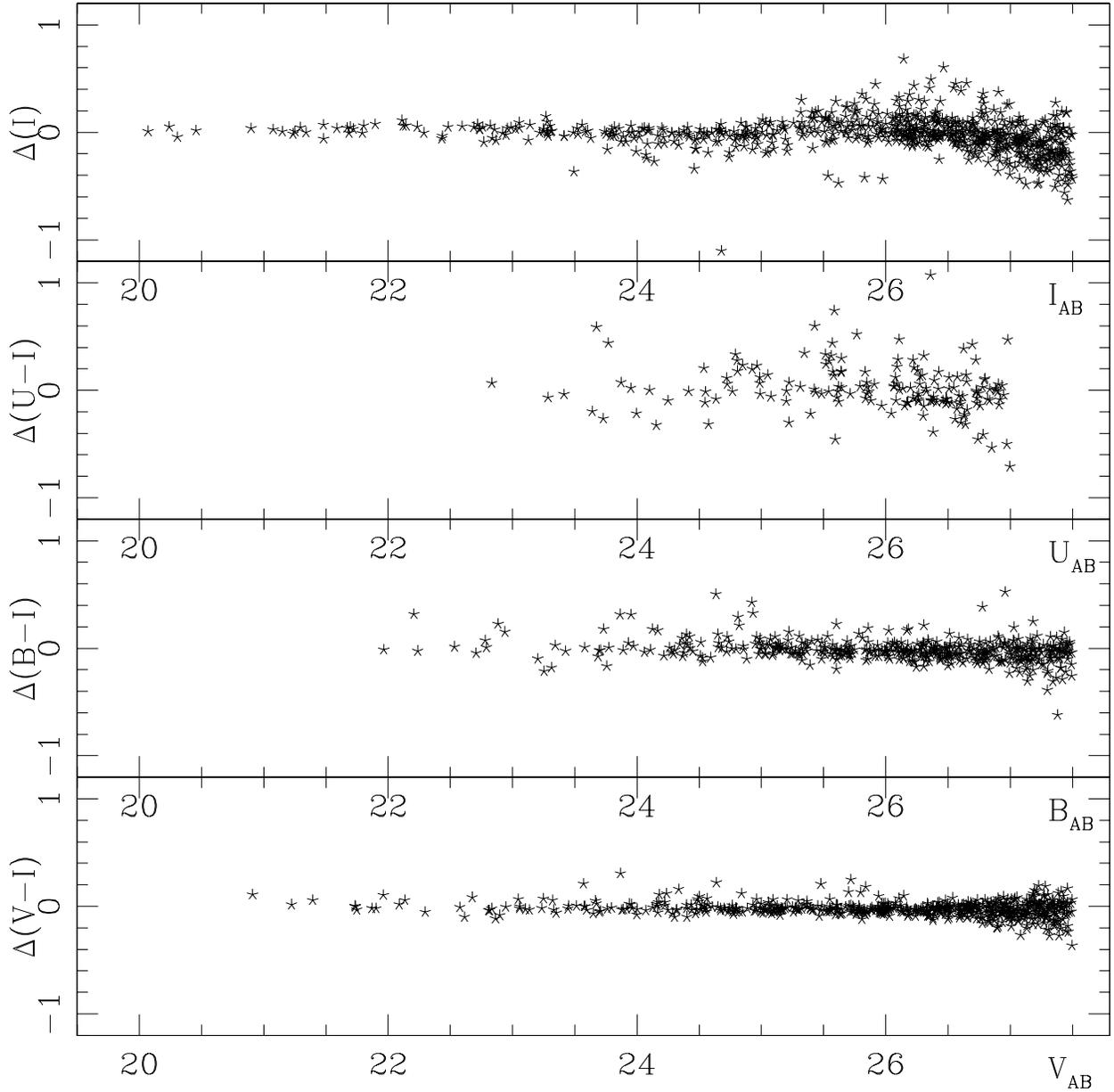}
 \caption{Comparison of the magnitudes and colors estimated in the present work
and the results of \citet{casertano00}. In abscissa the magnitudes,
in ordinate the differences between our estimates and those of 
\citet{casertano00}. \label{fig:Dcasert}}
  \end{figure}

 \begin{figure}
 \plotone{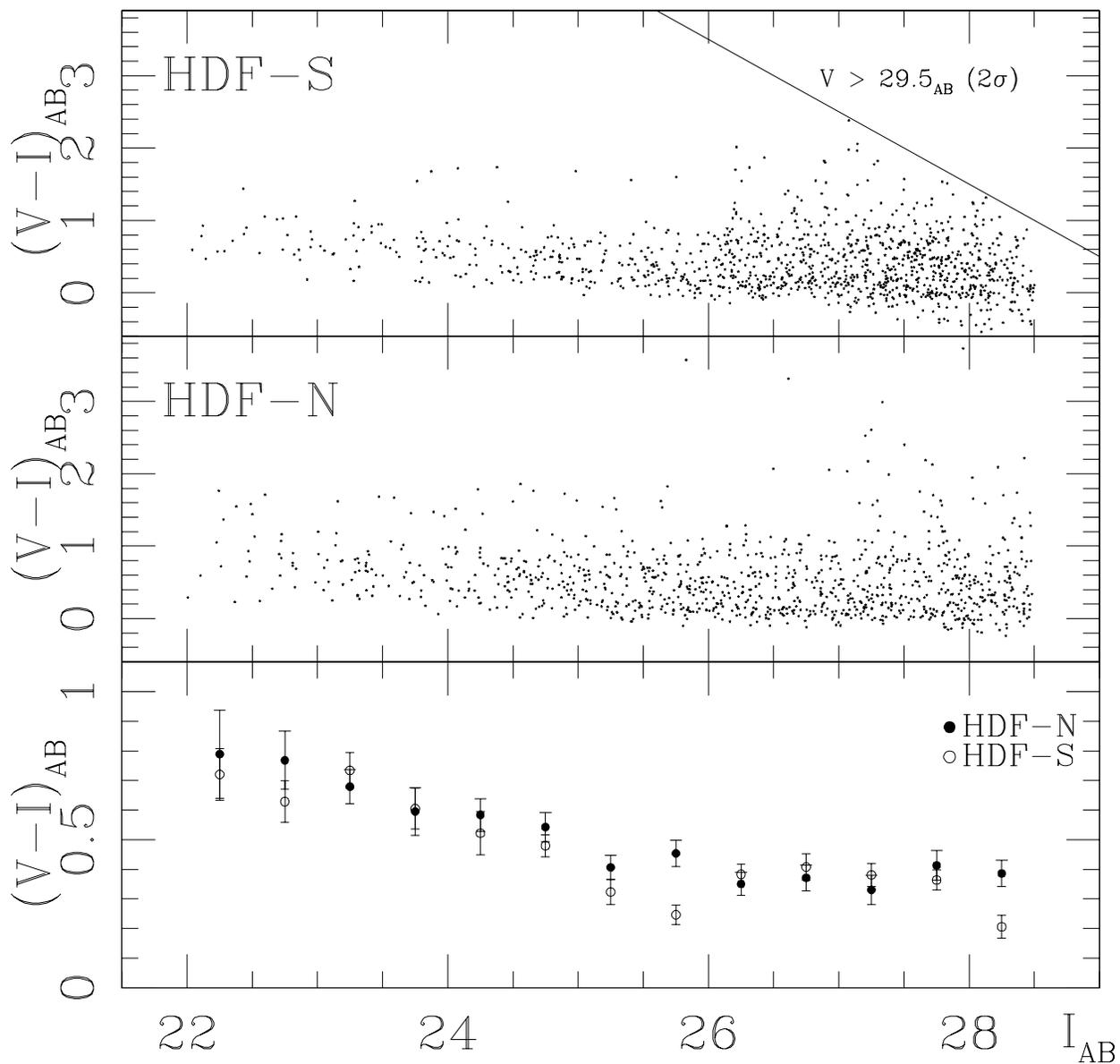}
 \caption{Comparison between HDF-S and HDF-N \citep{soto99}. 
In the first two panels the $V-I$ color versus $I$ (AB) is shown.  The
third panel shows the comparison between the median $V-I$ colors in
0.5 magnitude bins.  Beyond $I_{AB}$=28 the limited depth of the V
image (29.5, 2$\sigma$) gives origin to systematic effects on the
$V$-$I$ color.
\label{fig:colorV_I}}
  \end{figure}

  \begin{figure}
 \plotone{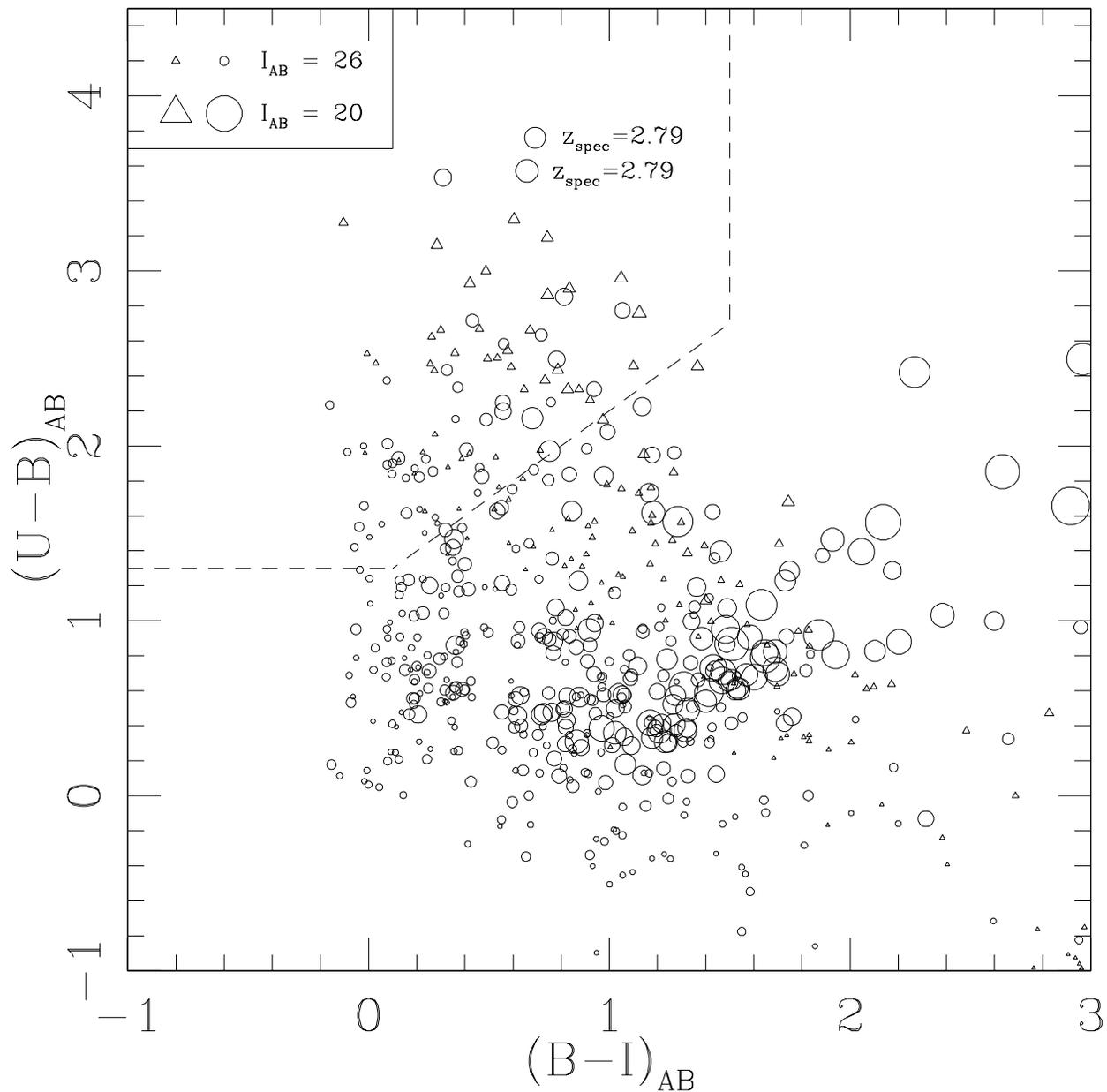}
 \caption{Color-color plot for the candidate U-band
 dropouts (above the dashed line) in the HDF-S. The size of the symbols scales
 with the magnitude in $I_{AB}$. There are two spectroscopic redshift
 available (from Cristiani et al. 2000). \label{fig:CandU}}
  \end{figure}

  \begin{figure}
  \plotone{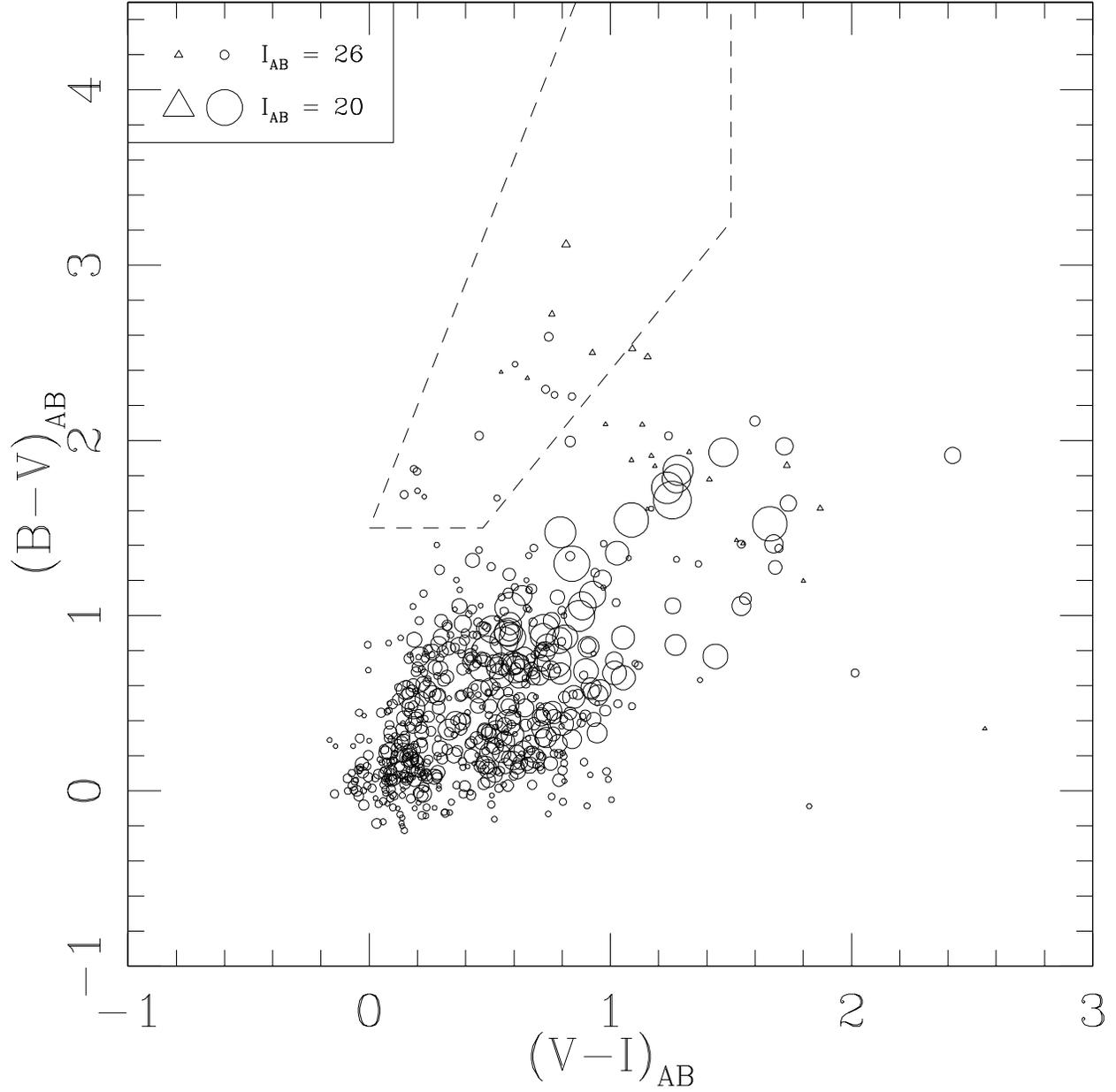}
 \caption{Color-color plot for the candidate B-band
 dropouts (above the dashed line) in the HDF-S. The size of the symbols scales
 with the magnitude in $I_{AB}$.
 \label{fig:CandB}}
  \end{figure}

 \begin{figure}
 \plotone{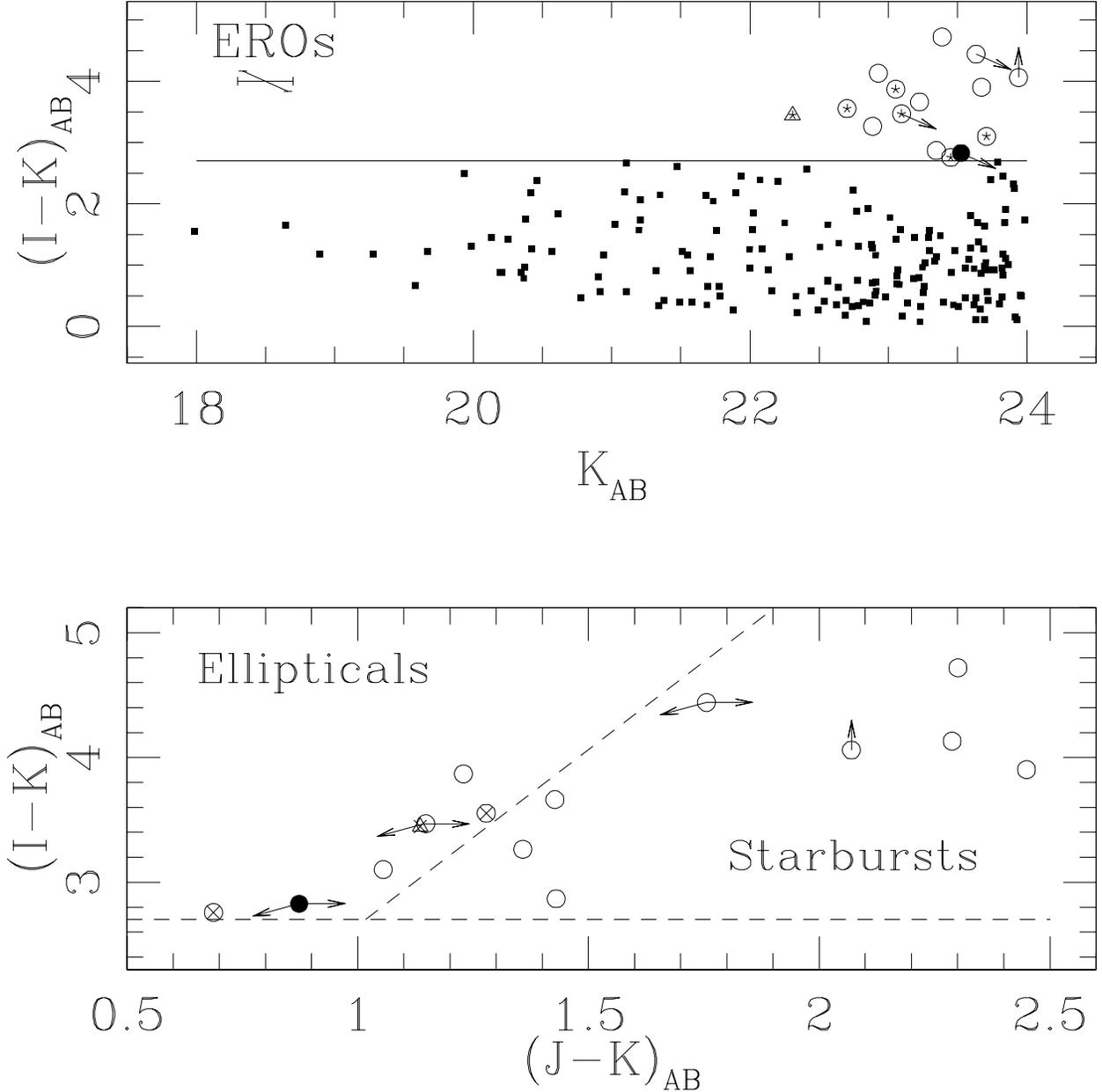}
 \caption{EROs in the HDF-S ($(I-K)_{AB}>2.7$ and
 $K_{AB}<24$). 
In the upper panel empty symbols indicate the $Starburst$ galaxies and
the stars the $Ellipticals$. The triangle corresponds to a
radio-galaxy (see text). In the upper panel the error bars are
plotted in the upper left corner for an object with $K_{AB}$ = 24 and
$(I-K)_{AB}$ = 3.5.
In the lower panel the classification between
$Elliptical$ and $Starburst$ is carried out with a color-color technique
(Pozzetti et al. 2000). Crosses indicate objects morphologically
classified as elliptical galaxies. The filled circle in both panels is an object
morphologically classified as ``late-type''.
\label{fig:CandEROs}}
 \end{figure}

\begin{figure}
\plotone{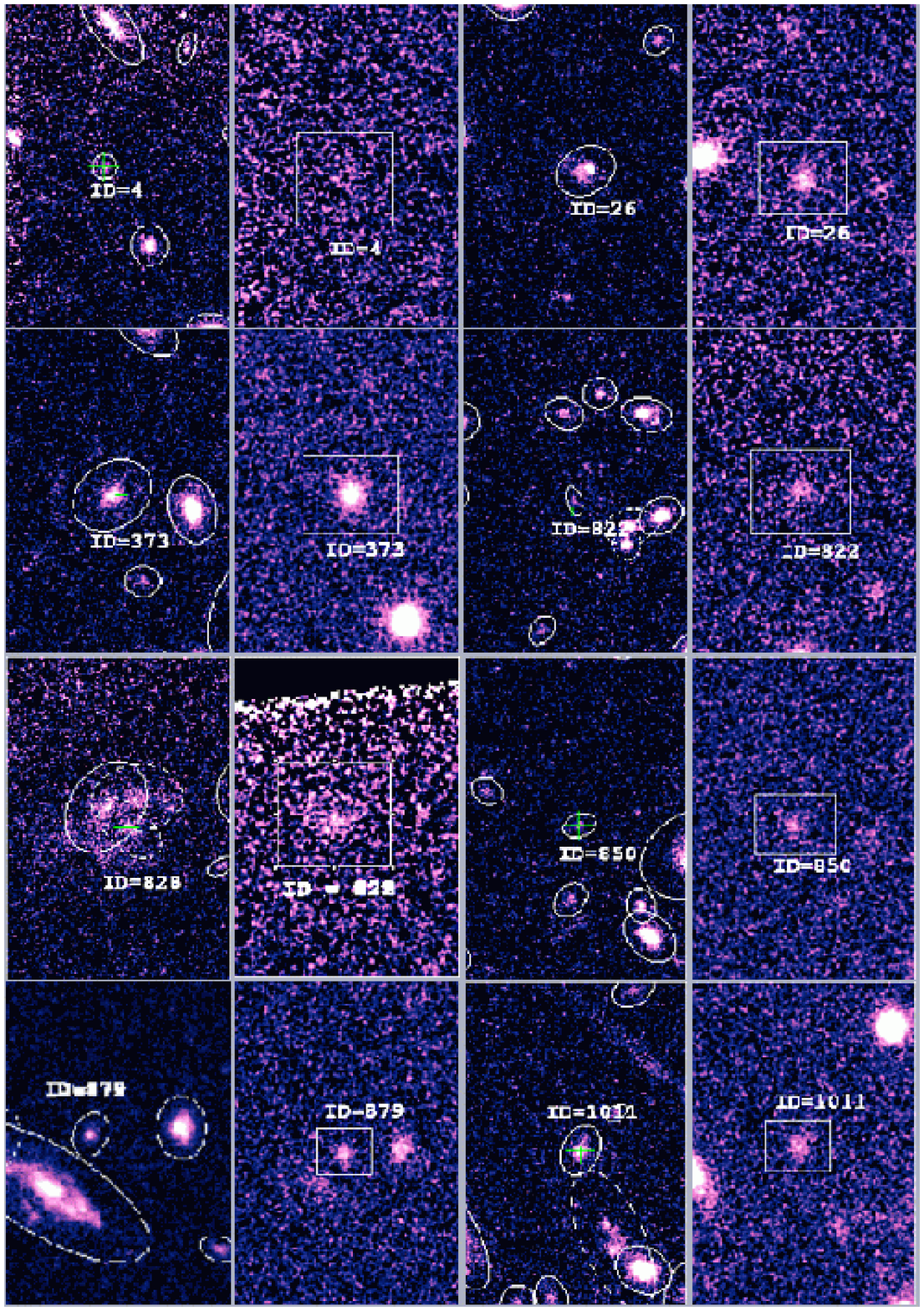}
\caption[vanzella.FIG10o.ps] {EROs in the $V+I$ and $Ks$ images, they are marked
with a cross and the identifier.
 \label{fig:EROsPOSTER1}}
\end{figure}

\begin{figure}
\plotone{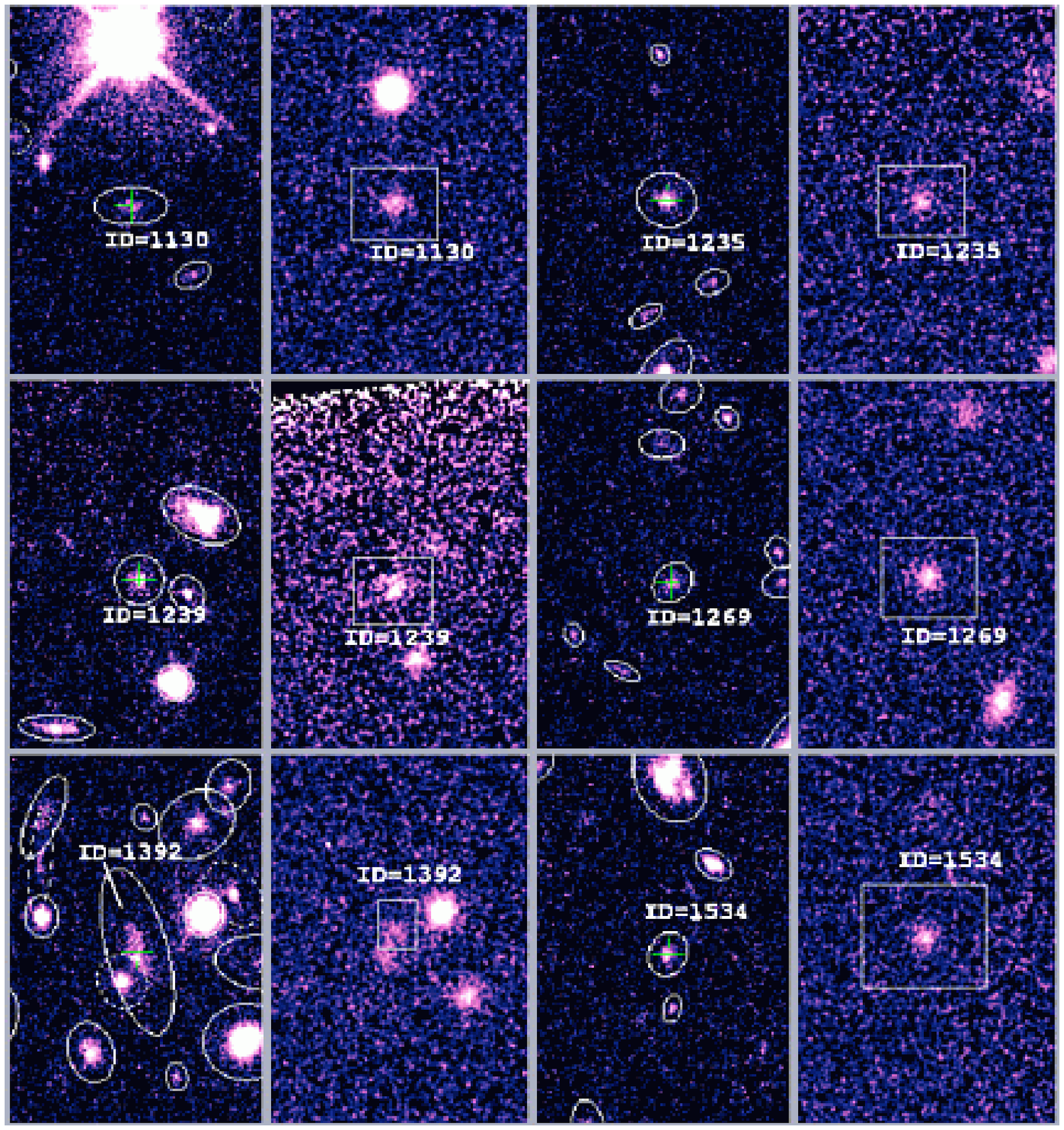}
 \caption[vanzella.FIG11k.ps] {EROs in the $V+I$ and $Ks$ images, they are
 marked with a cross and the identifier.
\label{fig:EROsPOSTER2}}
\end{figure}

%% file: vanzella.tab.tex
\clearpage
\begin{table}[t]  
\caption {WFPC2 observations}  
\begin{center}  
\label{table:tobs} 
\begin{tabular}{lcccccc} 
\tableline \tableline  
 Filter   &  Number of & total  &  FWHM   & Zero-point & Limiting AB
Magnitude\\  
          &   frames &  time (s) & (arcsec)  &   ABMAG  & (10-$\sigma$
in 0.2 square arcsec)\\   
\tableline   
F300     & 102 & 132585 & 0.14 & 20.77 & 26.8 \\ 
F450     & 51 & 101800 & 0.14 & 21.9 & 27.7 \\ 
F606     & 49 & 97200 & 0.14 & 23.02 & 28.3 \\ 
F814     & 56 & 112200 & 0.14 & 22.09 & 27.7 \\ 
\tableline  
\end{tabular}  
\end{center}  
\tablecomments{From Casertano et al. 2000.}  
\end{table} 

\begin{table} 
\caption {ISAAC observations} 
\begin{center} 
\label{table:tobsIR}
\begin{tabular}{lcccccccc}
\tableline \tableline 
 Filter   & $<\lambda>$ & $\Delta \lambda$ &Number of & total  &  FWHM
& Zero-point & Limiting Magnitude \\ 
          &           &                &    frames &  time (s) &
(arcsec)  &   & 5-$\sigma$ \\   
\tableline  
Js    & 1.24 & 0.16 & 210 & 25200 & 0.60 & 28.35 & 23.5\\
H     & 1.65 & 0.30 & 181 & 21720 & 0.60 & 27.50 & 22.0\\
Ks    & 2.16 & 0.27 & 487 & 29220 & 0.60 & 26.26 & 22.0\\
\tableline 
\end{tabular} 
\end{center} 
\tablecomments{Magnitudes are provided in the Vega magnitude system. 
} 
\end{table}

\begin{deluxetable}{rrrccrrrrrrrrrrrrrrrrl} 
\setlength{\tabcolsep}{0.025in}  
\rotate 
\tabletypesize {\scriptsize} 
\tablecolumns{22}
\tablewidth{0pc}  
\renewcommand{\arraystretch}{.7} 
 \tablecaption{The object catalog for the HDF-S WFPC2 field} 

\tablehead{  
\colhead{ID}&  
\colhead{x}& 
\colhead{y}&
\colhead{$\alpha(22^h:m:s)$}&
\colhead{$\delta(-60^{\circ}:':'')$}&
\colhead{$m_i$}& 
\colhead{$\sigma(m_i)$}& 
\colhead{Flux $U$}&
\colhead{$errU$}&
\colhead{Flux $B$}&
\colhead{$errB$}&
\colhead{Flux $V$}&
\colhead{$errV$}&
\colhead{Flux $J$}&
\colhead{$errJ$}&
\colhead{Flux $H$}&
\colhead{$errH$}&
\colhead{Flux $K$}&
\colhead{$errK$}&
\colhead{s/g}&
\colhead{$f_{s}$}&
\colhead{$f_{we}$}
}  
\startdata  
1 & 222.56 & 1849.99 & 33 05.771 & 33 23.32 & 27.08 & 0.11 & 0.0144 & 0.0079 & 0.0312 & 
0.0039 & 0.0382 & 0.0024& 0.3045 & 0.2701 & 0.1359
 & 0.4970 & 0.2725 & 0.4323 & 0.01 & 0 & 0 \\
2 & 229.87 & 1667.82 & 33 05.728 & 33 30.57 & 29.72 & 1.29 & 0.0009 & 0.0091 & -0.0032 & 
0.0041 & 0.0071 & 0.0024 & 0.1733 & 0.1826 & -0.0959 & 0.3911 & -0.1325 & 0.3180 & 0.53 & 0 & 7 \\
3 & 245.78 & 1812.88 & 33 05.647 & 33 24.79 & 27.75 & 0.19 & 0.0210 & 0.0076 & 
0.0131 & 0.0038 & 0.0140 & 0.0023& -0.0406 & 0.2279 & 0.2266 & 0.4699 & 0.3363 & 0.4302 & 0.20 & 0 & 0 \\
4 & 248.01 & 2109.94 & 33 05.647 & 33 12.95 & 28.08 & 0.34 & 0.0080 & 0.0096 & 
0.0015 & 0.0046 & 0.0153 & 0.0027 & -0.0589 & 0.2563 & 0.5490 
& 0.8582 & 0.9603 & 1.2918 & 0.88 & 0 & 0 \\
5 & 252.54 & 2205.65 & 33 05.625 & 33 09.13 & 26.00 & 0.09 & 0.0149 & 0.0126 & 0.0727 
& 0.0065 & 0.0903 & 0.0035 & 0.2667 & 0.1143 & 0.4916 
& 0.2870 & -0.1419 & 0.1902 & 0.03 & 0 & 0 \\
\enddata
\label{table:catalog}
\tablecomments{Fluxes and corresponding errors are in units 
of $10^{10} \rm{erg~cm^{-2}~s^{-1}~Hz^{-1}}$.\\
The complete version of this table is in the electronic
edition of the Journal.  The printed edition contains only a sample.}
\end{deluxetable}


\begin{table} 
\caption {Counts in the F814W band measured in the inner 
area of the WF camera ($4.49 $ arcmin$^{2}$ in the present work, 
$4.38 $ arcmin$^{2}$ in the work of Volonteri et al. (2000)).} 
\begin{center} 
\label{table:Counts}
\begin{tabular}{lccccccccc}
\tableline \tableline 
 $I_{AB}$ & $n$ & $c_{gal}$ & $c_{star}$& $Nc_{gal}$             & $Nc_{star}$           & $n_{vol}$ & $c_{vol}$ & $Nc_{vol}$ \\
          &     &           &           & {\small ($0.5~mag^{-1}$} & 
\small{($0.5~mag^{-1}$} &           &           & 
\small{($0.5~mag^{-1}$} \\ 
          &     &           &           & {\small $arcmin^{-2}$)} & 
\small{$arcmin^{-2}$)} &           &           & 
\small{$arcmin^{-2}$)} \\ 
\tableline 
22.25 & 11  & 1.04 & 1.04 &  2.55 &  2.55 &  11 & 1 & 2.51\\
22.75 & 17  & 1.04 & 1.04 &  3.94 &  3.94 &  28 & 1 & 6.39\\
23.25 & 25  & 1.04 & 1.04 &  5.79 &  5.79 &  26 & 1 & 5.94\\
23.75 & 28  & 1.04 & 1.04 &  6.49 &  6.49 &  35 & 1 & 7.99\\
24.25 & 35  & 1.04 & 1.04 &  8.11 &  8.11 &  37 & 1 & 8.45\\
24.75 & 53  & 1.04 & 1.04 & 12.28 & 12.28 &  65 & 1 & 14.84\\
25.25 & 46  & 1.05 & 1.05 & 10.76 & 10.76 &  94 & 1 & 21.46\\
25.75 & 79  & 1.07 & 1.06 & 18.83 & 18.65 & 139 & 1 & 31.73\\
26.25 & 143 & 1.11 & 1.06 & 35.35 & 33.76 & 173 & 1 & 39.50\\
26.75 & 157 & 1.23 & 1.07 & 43.01 & 37.41 & 235 & 1 & 53.65\\
27.25 & 205 & 1.59 & 1.08 & 72.59 & 49.31 & 342 & 1.08 & 84.33\\
27.75 & 222 & 2.87 & 1.12 & 141.90& 55.38 & 332 & 1.86 & 140.99\\
\tableline 
\end{tabular} 
\end{center} 
\tablecomments{$n$ and $n_{vol}$ are the observed counts in 0.5 magnitude
bins in the present work and in the catalog by Volonteri et al. (2000),
respectively. $c_{gal}$ and $c_{star}$ are the incompleteness correction
factors derived from our simulations for galaxies and
stars, respectively (see text). 
$Nc_{gal}$ and $Nc_{star}$ are the corrected counts derived in the
present work adopting the incompleteness correction estimated for
galaxies and for stars, respectively.
$c_{vol}$ is the incompleteness correction factor and $Nc_{vol}$ the
corrected counts per square arcmin as estimated by Volonteri et
al. (2000).} 
\end{table}

\begin{deluxetable}{rrrrrrrrl} 
\setlength{\tabcolsep}{0.12in}  
\tablecolumns{8}
\renewcommand{\arraystretch}{1.} 
 \tablecaption{HDF-S U-dropouts} 

\tablehead{  
\colhead{ID}&  
\colhead{I$_{AB}$}& 
\colhead{(B-I)$_{AB}$}&
\colhead{(U-B)$_{AB}$}& 
\colhead{X}&
\colhead{Y}& 
\colhead{RA}& 
\colhead{DEC}&
}  
\startdata  
28 & 24.20 & 0.31 & 3.53 & 381.90 & 1663.84 & 22 33 04.90 & -60 33 30.7 \\ 
46 & 26.26 & 0.26 & 2.47 & 445.65 & 1420.50 & 22 33 04.55 & -60 33 40.4 \\ 
49 & 26.44 & -0.01 & 2.53 & 456.64 & 1695.53 & 22 33 04.50 & -60 33 29.4 \\ 
56 & 24.88 & 0.83 & 2.90 & 479.55 & 1811.28 & 22 33 04.38 & -60 33 24.8 \\ 
57 & 26.59 & 0.42 & 1.96 & 489.72 & 1669.49 & 22 33 04.32 & -60 33 30.4 \\ 
80 & 25.34 & 0.73 & 2.37 & 565.62 & 1027.16 & 22 33 03.88 & -60 33 56.0 \\ 
81 & 24.30 & 0.56 & 2.20 & 565.77 & 2083.53 & 22 33 03.93 & -60 33 13.9 \\ 
89 & 25.90 & -0.04 & 1.54 & 596.78 & 1023.16 & 22 33 03.71 & -60 33 56.1 \\ 
109 & 25.09 & 0.28 & 3.15 & 669.68 & 571.08 & 22 33 03.30 & -60 34 14.1 \\ 
115 & 23.25 & 0.68 & 2.16 & 695.55 & 1685.70 & 22 33 03.21 & -60 33 29.7 \\ 
124 & 25.41 & 0.58 & 2.54 & 727.14 & 1083.73 & 22 33 03.01 & -60 33 53.7 \\ 
127 & 25.46 & 1.10 & 2.46 & 736.19 & 965.96 & 22 33 02.96 & -60 33 58.4 \\ 
219 & 24.74 & 0.47 & 1.83 & 1137.10 & 1426.53 & 22 33 00.81 & -60 33 39.9 \\ 
225 & 25.98 & 0.76 & 2.25 & 1152.38 & 872.99 & 22 33 00.70 & -60 34 02.0 \\ 
307 & 25.69 & 0.21 & 1.82 & 1402.15 & 1216.71 & 22 32 59.37 & -60 33 48.2 \\ 
329 & 26.16 & -0.16 & 2.23 & 1457.20 & 1077.04 & 22 32 59.06 & -60 33 53.7 \\ 
382 & 25.98 & 0.54 & 2.50 & 1567.01 & 1102.02 & 22 32 58.47 & -60 33 52.7 \\ 
385 & 25.54 & 0.49 & 3.00 & 1570.16 & 1127.06 & 22 32 58.46 & -60 33 51.7 \\ 
396 & 24.74 & 0.74 & 2.86 & 1604.91 & 1695.32 & 22 32 58.29 & -60 33 29.0 \\ 
536 & 26.08 & 0.10 & 1.90 & 1986.10 & 907.38 & 22 32 56.20 & -60 34 00.3 \\ 
540 & 26.05 & -0.02 & 1.66 & 1989.17 & 620.40 & 22 32 56.17 & -60 34 11.7 \\ 
583 & 24.87 & 0.60 & 3.29 & 2072.45 & 3106.80 & 22 32 55.83 & -60 32 32.6 \\ 
600 & 26.72 & 0.33 & 1.88 & 2124.12 & 4053.40 & 22 32 55.60 & -60 31 54.8 \\ 
638 & 26.09 & 0.46 & 1.88 & 2213.60 & 1214.37 & 22 32 54.98 & -60 33 48.0 \\ 
646 & 26.38 & 0.36 & 2.15 & 2229.45 & 1710.16 & 22 32 54.92 & -60 33 28.2 \\ 
649 & 25.70 & -0.10 & 3.28 & 2234.97 & 1359.88 & 22 32 54.87 & -60 33 42.2 \\ 
663 & 26.74 & 0.05 & 1.55 & 2268.45 & 1293.80 & 22 32 54.69 & -60 33 44.8 \\ 
676 & 26.11 & 0.24 & 1.93 & 2290.75 & 2716.97 & 22 32 54.63 & -60 32 48.1 \\ 
685 & 25.80 & 0.87 & 2.32 & 2308.56 & 2887.93 & 22 32 54.54 & -60 32 41.2 \\ 
693 & 26.34 & -0.06 & 1.42 & 2322.93 & 699.27 & 22 32 54.37 & -60 34 08.5 \\ 
709 & 26.82 & 0.23 & 1.63 & 2341.94 & 2472.97 & 22 32 54.35 & -60 32 57.8 \\ 
719 & 25.56 & 0.08 & 2.01 & 2354.33 & 2476.44 & 22 32 54.28 & -60 32 57.6 \\ 
723 & 24.56 & 0.56 & 2.25 & 2359.55 & 682.90 & 22 32 54.16 & -60 34 09.1 \\ 
738 & 25.30 & 0.72 & 2.63 & 2387.22 & 3168.21 & 22 32 54.13 & -60 32 30.1 \\ 
757 & 26.57 & 0.39 & 1.93 & 2415.28 & 904.51 & 22 32 53.88 & -60 34 00.3 \\ 
776 & 25.85 & 0.46 & 2.67 & 2447.00 & 4070.66 & 22 32 53.85 & -60 31 54.0 \\ 
778 & 26.22 & 0.10 & 1.84 & 2450.30 & 967.91 & 22 32 53.69 & -60 33 57.7 \\ 
784 & 26.31 & 0.71 & 1.98 & 2455.43 & 3453.57 & 22 32 53.78 & -60 32 18.7 \\ 
802 & 24.54 & 1.05 & 2.77 & 2507.67 & 2110.67 & 22 32 53.43 & -60 33 12.2 \\ 
807 & 24.67 & 0.94 & 2.32 & 2512.53 & 3856.67 & 22 32 53.49 & -60 32 02.5 \\ 
829 & 25.03 & 0.79 & 2.44 & 2544.12 & 1906.26 & 22 32 53.23 & -60 33 20.3 \\ 
847 & 26.60 & 0.28 & 1.59 & 2571.80 & 4045.39 & 22 32 53.18 & -60 31 55.0 \\ 
855 & 26.41 & 0.03 & 2.47 & 2582.79 & 4120.85 & 22 32 53.12 & -60 31 52.0 \\ 
868 & 26.94 & -0.01 & 1.96 & 2606.04 & 1104.37 & 22 32 52.85 & -60 33 52.2 \\ 
870 & 26.33 & 0.15 & 1.82 & 2606.41 & 1228.31 & 22 32 52.86 & -60 33 47.3 \\ 
871 & 24.89 & 0.74 & 3.19 & 2606.45 & 1869.42 & 22 32 52.89 & -60 33 21.7 \\ 
913 & 26.92 & 0.13 & 1.92 & 2676.51 & 2941.20 & 22 32 52.56 & -60 32 39.0 \\ 
914 & 25.62 & 0.56 & 2.58 & 2681.97 & 2405.91 & 22 32 52.51 & -60 33 00.3 \\ 
949 & 25.26 & 0.49 & 2.15 & 2740.41 & 2984.83 & 22 32 52.21 & -60 32 37.2 \\ 
957 & 24.66 & 1.05 & 2.96 & 2759.70 & 1344.73 & 22 32 52.03 & -60 33 42.6 \\ 
960 & 26.62 & -0.02 & 2.00 & 2765.76 & 2729.73 & 22 32 52.07 & -60 32 47.4 \\ 
968 & 26.68 & 0.23 & 1.96 & 2778.16 & 4205.43 & 22 32 52.08 & -60 31 48.6 \\ 
991 & 26.69 & 0.54 & 1.76 & 2808.66 & 3222.52 & 22 32 51.86 & -60 32 27.7 \\ 
992 & 24.48 & 1.12 & 2.76 & 2808.81 & 3768.18 & 22 32 51.89 & -60 32 06.0 \\ 
1007 & 26.54 & 0.53 & 1.94 & 2834.69 & 3331.53 & 22 32 51.72 & -60 32 23.4 \\ 
1025 & 25.06 & 0.42 & 2.93 & 2874.27 & 2037.31 & 22 32 51.45 & -60 33 14.9 \\ 
1068 & 26.94 & 0.23 & 1.50 & 2938.20 & 3806.90 & 22 32 51.19 & -60 32 04.4 \\ 
1081 & 26.99 & 0.38 & 1.64 & 2964.56 & 3705.72 & 22 32 51.04 & -60 32 08.4 \\ 
1100 & 23.32 & 0.75 & 1.97 & 3012.83 & 1694.70 & 22 32 50.68 & -60 33 28.6 \\ 
1123 & 26.43 & 0.19 & 1.87 & 3052.65 & 3719.02 & 22 32 50.57 & -60 32 07.9 \\ 
1149 & 25.91 & 0.50 & 2.50 & 3091.54 & 2122.05 & 22 32 50.28 & -60 33 11.5 \\ 
1167 & 25.93 & 0.30 & 2.66 & 3121.38 & 1434.79 & 22 32 50.08 & -60 33 38.9 \\ 
1174 & 24.14 & 0.81 & 2.85 & 3125.97 & 985.42 & 22 32 50.04 & -60 33 56.8 \\ 
1179 & 25.84 & 0.27 & 1.85 & 3137.28 & 3750.66 & 22 32 50.11 & -60 32 06.6 \\ 
1196 & 26.90 & 0.19 & 1.84 & 3175.93 & 1894.17 & 22 32 49.81 & -60 33 20.5 \\ 
1204 & 26.42 & -0.09 & 1.96 & 3197.90 & 1817.71 & 22 32 49.69 & -60 33 23.6 \\ 
1212 & 25.04 & 0.12 & 1.93 & 3211.36 & 2333.12 & 22 32 49.64 & -60 33 03.0 \\ 
1225 & 25.54 & 0.67 & 2.66 & 3234.36 & 826.16 & 22 32 49.45 & -60 34 03.1 \\ 
1232 & 25.69 & 0.37 & 2.33 & 3256.86 & 3861.56 & 22 32 49.47 & -60 32 02.1 \\ 
1242 & 22.92 & 0.66 & 3.57 & 3265.10 & 3254.94 & 22 32 49.40 & -60 32 26.3 \\ 
1249 & 26.24 & 0.27 & 2.43 & 3271.88 & 1251.97 & 22 32 49.26 & -60 33 46.1 \\ 
1261 & 25.90 & 0.59 & 2.45 & 3288.07 & 2844.31 & 22 32 49.26 & -60 32 42.6 \\ 
1277 & 23.28 & 0.69 & 3.76 & 3305.11 & 3261.82 & 22 32 49.18 & -60 32 26.0 \\ 
1281 & 26.86 & 0.00 & 1.48 & 3316.19 & 2780.19 & 22 32 49.09 & -60 32 45.2 \\ 
1303 & 25.13 & 0.43 & 2.72 & 3350.99 & 3022.72 & 22 32 48.92 & -60 32 35.5 \\ 
1322 & 25.52 & 0.16 & 1.62 & 3401.43 & 2983.13 & 22 32 48.65 & -60 32 37.0 \\ 
1372 & 25.79 & 0.92 & 2.26 & 3459.33 & 3083.83 & 22 32 48.34 & -60 32 33.0 \\ 
1374 & 26.44 & 0.45 & 1.73 & 3469.88 & 3413.81 & 22 32 48.30 & -60 32 19.9 \\ 
1397 & 24.86 & 0.83 & 2.32 & 3508.06 & 2171.68 & 22 32 48.03 & -60 33 09.4 \\ 
1411 & 26.56 & 0.28 & 2.07 & 3529.57 & 1963.53 & 22 32 47.90 & -60 33 17.6 \\ 
1414 & 26.69 & 0.21 & 1.64 & 3533.03 & 3549.59 & 22 32 47.97 & -60 32 14.4 \\ 
1432 & 24.22 & 0.78 & 2.49 & 3562.83 & 3420.07 & 22 32 47.80 & -60 32 19.6 \\ 
1452 & 25.94 & 0.65 & 2.32 & 3589.04 & 3993.43 & 22 32 47.68 & -60 31 56.7 \\ 
1457 & 26.02 & 0.26 & 2.62 & 3599.39 & 1368.64 & 22 32 47.50 & -60 33 41.3 \\ 
1466 & 25.86 & 0.36 & 2.53 & 3618.27 & 2690.07 & 22 32 47.46 & -60 32 48.6 \\ 
1523 & 25.51 & 0.33 & 2.43 & 3721.70 & 2488.36 & 22 32 46.89 & -60 32 56.7 \\ 
1524 & 26.09 & 0.08 & 1.89 & 3725.12 & 2092.66 & 22 32 46.85 & -60 33 12.4 \\ 
1532 & 26.38 & 0.08 & 2.37 & 3740.41 & 2895.03 & 22 32 46.81 & -60 32 40.4 \\ 
1572 & 26.92 & 0.29 & 1.56 & 3830.48 & 1300.51 & 22 32 46.25 & -60 33 44.0 \\ 
1588 & 24.95 & 0.41 & 1.98 & 3883.97 & 1288.26 & 22 32 45.95 & -60 33 44.4 \\ 
\enddata
\label{table:Udropouts}
\tablecomments{List of U-band dropout candidates.} 
\end{deluxetable}

\begin{deluxetable}{rrrrrrrrl} 
\setlength{\tabcolsep}{0.12in}  
\tablecolumns{8}
\renewcommand{\arraystretch}{1.} 
 \tablecaption{HDF-S B-dropouts} 

\tablehead{  
\colhead{ID}&  
\colhead{I$_{AB}$}& 
\colhead{(V-I)$_{AB}$}&
\colhead{(B-V)$_{AB}$}& 
\colhead{X}&
\colhead{Y}& 
\colhead{RA}& 
\colhead{DEC}&
}  
\startdata  
36 & 26.79 & 0.20 & 1.71 & 397.39 & 1030.43 & 22 33 04.79 & -60 33 55.9 \\ 
209 & 26.38 & 0.84 & 2.25 & 1119.71 & 815.73 & 22 33 00.88 & -60 34 04.2 \\ 
320 & 26.65 & 0.53 & 1.67 & 1425.84 & 1247.15 & 22 32 59.24 & -60 33 46.9 \\ 
344 & 26.47 & 0.19 & 1.84 & 1479.11 & 917.34 & 22 32 58.94 & -60 34 00.1 \\ 
369 & 26.08 & 0.74 & 2.59 & 1539.58 & 692.30 & 22 32 58.60 & -60 34 09.0 \\ 
442 & 26.31 & 0.20 & 1.82 & 1745.15 & 1205.75 & 22 32 57.51 & -60 33 48.5 \\ 
627 & 26.97 & 0.23 & 1.68 & 2196.19 & 2326.58 & 22 32 55.13 & -60 33 03.6 \\ 
721 & 26.89 & 0.66 & 2.36 & 2358.17 & 2817.65 & 22 32 54.27 & -60 32 44.0 \\ 
892 & 25.77 & 0.82 & 3.12 & 2653.70 & 2543.59 & 22 32 52.67 & -60 32 54.8 \\ 
897 & 27.00 & 0.55 & 2.39 & 2659.16 & 3657.16 & 22 32 52.69 & -60 32 10.5 \\ 
1004 & 26.22 & 0.73 & 2.29 & 2831.33 & 2109.59 & 22 32 51.69 & -60 33 12.1 \\ 
1144 & 26.43 & 0.93 & 2.50 & 3084.84 & 2152.57 & 22 32 50.32 & -60 33 10.3 \\ 
1151 & 26.79 & 0.60 & 2.43 & 3093.98 & 2959.69 & 22 32 50.30 & -60 32 38.1 \\ 
1155 & 26.23 & 0.14 & 1.69 & 3099.80 & 2050.02 & 22 32 50.23 & -60 33 14.4 \\ 
1400 & 26.53 & 0.77 & 2.26 & 3516.67 & 3877.74 & 22 32 48.07 & -60 32 01.4 \\ 
1461 & 26.08 & 0.46 & 2.03 & 3606.20 & 3451.91 & 22 32 47.56 & -60 32 18.3 \\ 
1494 & 26.38 & 0.76 & 2.72 & 3670.06 & 3641.84 & 22 32 47.23 & -60 32 10.7 \\  
\enddata
\label{table:Bdropouts}
\tablecomments{List of B-band dropout candidates.} 
\end{deluxetable}

\begin{table} 
\caption {EROs} 
\begin{center} 
\label{table:TABEROs}
\begin{tabular}{lccccccc}
\tableline \tableline 
 ID   & RA & DEC & $K_{AB}$ & $I_{AB}$ & $(I-K)_{AB}$ & Morphology \\

\tableline  
   4 & 22 33 05.65 & -60 33 12.90 &  23.63 &  28.08  &  4.44* & $S$\\    
  26 & 22 33 04.96 & -60 33 22.20 &  22.89 &  26.15  &  3.26  & $S$\\   
 373 & 22 32 58.60 & -60 33 46.58 &  22.31 &  25.75  &  3.45  & $E (Radio~Gal.)$  \\
 822 & 22 32 53.39 & -60 31 54.33 &  23.39 &  28.11  &  4.72  & $S$ \\
 828 & 22 32 53.34 & -60 31 46.58 &  23.09 &  26.56  &  3.47* & $S$   \\ 
 850 & 22 32 53.01 & -60 33 57.03 &  23.67 &  27.58  &  3.90  & $S$ \\  
 879 & 22 32 52.91 & -60 32 15.72 &  23.71 &  26.81  &  3.10  & $S$  \\ 
1011 & 22 32 51.54 & -60 33 58.40 &  23.35 &  26.21  &  2.87  & $S$   \\ 
1130 & 22 32 50.49 & -60 32 22.64 &  23.23 &  26.89  &  3.66  & $S$  \\  
1235 & 22 32 49.46 & -60 31 57.88 &  23.45 &  26.21  &  2.76  & $E$  \\
1239 & 22 32 49.46 & -60 31 46.58 &  22.70 &  26.25  &  3.55  & $E$   \\
1269 & 22 32 49.25 & -60 32 11.62 &  22.93 &  27.06  &  4.13  & $S$  \\ 
1392 & 22 32 48.15 & -60 32 17.47 &  23.52 &  26.35  &  2.83* & $L$   \\
1534 & 22 32 46.79 & -60 32 33.74 &  23.05 &  26.92  &  3.87  & $S$ \\ 
9999 & 22 32 55.92 & -60 32 50.28 &  23.95 &   ---   &  $>4$  & $S$ \\
\tableline 
\end{tabular} 
\end{center} 
\tablecomments{A list of EROs in the HDF-S with $(I-K)_{AB} > 2.7$. 
The colors marked with an asterisk (*) are upper limits in the
infrared bands. The morphological classification is listed in Column 7
as: $E$ ``early-type'', 
$L$ ``late-type'', $S$ too faint for a reliable classification.
The one object 
detected with a S/N$>5$ in the $Ks$ band by Saracco et al. (2001) that
is not present in our multicolor catalog is listed with the ID 9999.} 
\end{table}